\documentclass[12pt,preprint]{aastex}
\usepackage{subfigure}
\begin{document}

\title{A Correlation Between Circumstellar Disks and Rotation in the Upper Scorpius OB Association}

\author{S. E. Dahm\altaffilmark{1}, Catherine L. Slesnick\altaffilmark{2,}\altaffilmark{3}, \& R. J. White\altaffilmark{4}}

\altaffiltext{1}{W. M. Keck Observatory, 65-1120 Mamalahoa Hwy, Kamuela, HI 96743}
\altaffiltext{2}{Department of Terrestrial Magnetism, Carnegie Institution for Science, Washington, DC 20015}
\altaffiltext{3}{Current address: Charles Stark Draper Laboratory, 555 Technology Sq MS 31, Cambridge MA 02139}
\altaffiltext{4}{Department of Physics and Astronomy, Georgia State University, Atlanta, GA 30303, USA}

\begin{abstract}
We present projected rotational velocities for 20 early-type (B8--A9) and 74 late-type (F2--M8)
members of the $\sim$5 Myr old Upper Scorpius OB Association derived from high dispersion
optical spectra obtained with the High Resolution Echelle Spectrometer (HIRES) on Keck I and 
the Magellan Inamori Kyocera Echelle (MIKE) on the Magellan Clay telescope. The spectroscopic 
sample is composed of stars and brown dwarfs with infrared signatures of circumstellar disks, both primordial 
and debris, and non-excess sources of comparable spectral type. We merge projected rotational 
velocities, accretion diagnostics, and {\it Spitzer Space Telescope} Infrared Array Camera 
(IRAC) and Multiband Imaging Photometer for {\it Spitzer} (MIPS) 24 $\mu$m photometry to examine 
the relationship between rotation and circumstellar disks. The rotational velocities are 
strongly correlated with spectral type, a proxy for mass, such that the median $v$sin$i$ for
B8--A9 type stars is: 195$\pm$70 km s$^{-1}$, F2--K4: 37.8$\pm$7.4 km s$^{-1}$, 
K5--K9: 13.8$^{+21.3}_{-8.2}$ km s$^{-1}$, M0--M5: 16.52$\pm$5.3 km s$^{-1}$, and 
M5.5--M8: 17.72$\pm$8.1 km s$^{-1}$. We find with a probability of $\ge$0.99 that M-type stars
and brown dwarfs having infrared excess suggestive of circumstellar disks rotate more slowly
than their non-excess counterparts. A similar correlation is present among F2--K9 type stars,
but only at the $\sim$97\% confidence level. Among the early-type (B8--A9) members, rotational 
velocities of the debris-disk and non-disk populations are indistinguishable. Considering the
late-type (F2--M8) stars and brown dwarfs, we find a low fraction of slowly rotating, non-excess 
sources relative to younger star forming regions, suggesting that most have spun up following 
disk dissipation. The few late-type (F2--M5)
debris disk sources, which may be representative of stars that have recently dispersed their
inner disks, are evenly divided between slow and moderate rotators.
\end{abstract}


\section{Introduction}
Understanding the evolution of angular momentum in the star formation process remains a fundamental 
theoretical problem despite decades of effort directed toward its resolution. Simply stated, the 
specific angular momentum of a typical molecular cloud core is several orders of magnitude greater 
than that of a solar-mass, zero-age main sequence (ZAMS) star (for a review see Bodenheimer 1995). 
While a combination of mechanisms is likely responsible for the angular momentum loss over the
entirety of pre-main sequence evolution, the near-ubiquitous presence of circumstellar disks around
protostars is suggestive of their critical role in the early regulation of angular momentum.

Stellar rotation rates are generally determined observationally using one of two techniques. Many 
early studies of rotation in pre-main sequence stars employed high-dispersion optical spectra to 
measure rotational line broadening (e.g. Vogel \& Kuhi 1981; Hartmann et al. 1986; Bouvier et al. 1986). 
The resulting projected rotational velocities ($v$sin$i$) are subject to uncertainties in the 
inclination of the stellar rotational axis relative to the line of sight, but can be determined 
from single observations and are not impacted by accretion-induced variability. Later investigations 
directly measured rotation periods of pre-main sequence stars by photometric monitoring (e.g. Herbst et al. 1986; 
Bouvier et al. 1993; Bouvier et al. 1997; Makidon et al. 2004; Lamm et al. 2004). These time-series 
observations are sensitive to active regions on the stellar surface and allow for the determination of 
rotation periods with an accuracy of $\sim$1\%. For accreting sources, however, variability complicates 
the determination of rotation periods, potentially biasing photometric surveys toward non-accreting sources.

Initial studies into the relationship between rotation and circumstellar disks for pre-main sequence 
stars found that T Tauri stars (TTS) are in general slow rotators with mean rotational velocities of 
$\sim$15 km s$^{-1}$ (Bodenheimer 1995 and references therein). The low rotational velocities imply 
that low-mass stars lose most of the specific angular momentum inherited from their parent molecular 
cloud well before they become optically detectable (Hartmann et al. 1986). Theoretical arguments given 
to explain the observed slow rotators generally invoke a manifestation of the disk locking mechanism 
whereby young pre-main sequence stars transfer angular momentum to their surrounding circumstellar 
disk (K\"{o}nigl 1991) or to an accretion-driven wind originating at the boundary between the disk 
and the stellar magnetosphere (Shu et al. 2000; Matt \& Pudritz 2005). In these
scenarios stellar rotation is regulated by the accretion disk. Therefore, once accretion stops and
the star has dissipated its inner disk, it should become `unlocked' and begin to spin
up as it contracts toward the main sequence. Observational evidence in regards to this disk-braking
paradigm, however, is conflicting.

Early investigations in favor of the classical disk-braking scenario (e.g. Edwards et al. 1993; 
Bouvier et al. 1993) that used photometrically derived rotation periods, found that accreting 
classical TTS (CTTS) have longer rotation periods on average than non-accreting, weak-line TTS (WTTS) in 
young ($\lesssim$ 3 Myr) star forming regions. The photometric monitoring campaigns of Attridge \& Herbst (1992), 
Choi \& Herbst (1996), and Herbst et al. (2002) identified a bimodal period distribution in the $\sim$1 Myr old 
Orion Nebula Cluster (ONC), which was interpreted as evidence for disk-braking among the slower rotators with 
masses $\ge$0.25 M$_{\odot}$. Stassun et al. (1999, 2001), however, challenged these results arguing that 
rotation periods in the ONC were not significantly different from being uniformly distributed. Lamm et al. (2005) 
provided additional support for the disk regulation scenario finding that slower rotators are more likely 
to show evidence of circumstellar disks in the $\sim$1--3 Myr old cluster NGC\,2264. Rebull (2001) and 
Makidon et al. (2004), however, find no clear correlation between rotation period and infrared excess in 
the ONC or NGC\,2264, respectively.

Rebull et al. (2006) attribute the conflicting results to two critical factors: (1) the use of
near-infrared (i.e. $JHK-$band) excesses that are insensitive to disks with inner radii extending 
beyond the dust sublimation limit, and (2) the need for large (i.e. several hundred stars) sample
sizes in order to distinguish between locked and freely spinning sources. To address these problems,
Rebull et al. (2006) use {\it Spitzer Space Telescope} ($Spitzer$; Werner et al. 2004) observations
of $\sim$500 sources in the ONC with photometrically-derived rotation periods to probe disk emission 
at larger orbital radii. The Rebull et al. (2006) study finds that nearly all stars with short periods
($\lesssim$2 days, corresponding to $<$$v$sin$i$$>$$\gtrsim$30 km/s for a typical TTS) are diskless, 
providing the most concrete evidence to date that the disk locking paradigm is correct. Among the
stars with long periods, about half are associated with disks. Rebull et al. (2006) 
conclude that of the stars without disks, those with long periods likely have dissipated their 
disks more recently than those stars with short periods which, presumably, have already had time
to spin-up. If this interpretation is correct, we would expect to see a much smaller fraction of
diskless stars rotating slowly in more evolved ($\sim$3--10 Myr-old) clusters and associations for
which the majority of stars dissipated their inner primordial disks many megayears ago.

Hartmann (2002) also discusses the need for studies of large clusters of intermediate-age ($\sim$3--10 Myr) pre-main 
sequence stars. This work re-examines the magnetospheric disk-braking scenario, arguing that the rate of stellar 
spin-down is limited by the rate at which angular momentum can be removed from the inner disk by viscous processes 
or by winds. Hartmann (2002) concludes that further measurements of rotation
periods as a function of age are neeeded to test the disk-braking hypothesis.

The Upper Scorpius OB association is a critically important region for studies of disk evolution. At $\sim$145 pc 
distant, it is among the nearest OB associations to the Sun (Blaauw 1991; de Zeeuw et al. 1999) and has a 
well-established age of $\sim$5 Myr (Preibisch \& Zinnecker 1999; Preibisch et al. 2002), when most ($\sim$80\%) 
optically thick, primordial disks have dissipated (Haisch et al. 2001; Carpenter et al. 2006; Hernandez et al. 2007; 
Dahm \& Hillenbrand 2007), but many optically thin debris disks remain. Of particular significance, the age dispersion 
within Upper Scorpius is estimated to be $\le$3 Myr (Preibisch et al. 2002; Slesnick et al. 2008). Furthermore, Upper Scorpius contains 
several hundred spectroscopically confirmed members across a large mass range ($\sim$15--0.02 M$_\odot$). 
Carpenter et al. (2006) conducted a {\it Spitzer} 4.5--16 $\mu$m photometric survey of 218 confirmed association 
members for infrared excess emission, identifying 35 stars with 8 or 16 $\mu$m excess. Only 19$^{+5}_{-4}$\% of 
K+M stars in Upper Scorpius were found with infrared excess emission indicative of primordial disks. The follow-up
24 and 70 $\mu$m photometric survey of Carpenter et al. (2009) identified 19 debris-like systems in Upper Scorpius 
that exhibit 24 or 70 $\mu$m excess, but lack excess emission at shorter wavelengths.

Conceptually, primordial disks retain significant quantities of gas and dust that are the remnants of 
the star formation process. Included within the primordial disk category are sub-classes including:
transition disks (Strom et al. 1989), pre-transitional disks (Espaillat et al. 2007), anemic disks 
(Lada et al. 2006), and homologously depleted disks (Currie et al. 2009), which are all characterized
by reduced levels of emission at wavelengths $\le$10 $\mu$m, but that resemble CTTS at longer wavelengths.
The 24 and 70 $\mu$m excesses of debris disk sources presumably originate from the collision of planetesimals
(i.e. second-generation dust), although the possibility that these are the remnants of primordial disks 
with substantial inner holes cannot be ruled out (Carpenter et al. 2009). The 24 $\mu$m excesses around
the late-type (K+M) debris disk systems included here appear to be distinct from transitional disk systems
(Carpenter et al. 2009).

In this work we present projected rotational velocities for a sample of 94 Upper Scorpius stellar and substellar 
members drawn from the {\it Spitzer} mid-infrared surveys of Carpenter et al. (2006) and Slesnick (2007) and 
the optical photometric and spectroscopic study of Slesnick et al. (2006; 2008). Given the spatial extent of the Upper 
Scorpius OB association ($>$100 square degrees), photometric rotation period surveys are more challenging than
for compact star forming regions like the ONC, NGC\,2264, and IC\,348. Adams et al. (1998) provide rotation 
periods for 3 Upper Scorpius members included here, ScoPMS 17: 2.93 days, ScoPMS 31: 1.32 days, and ScoPMS 45: 
5.81 days. More comprehensive rotational period surveys are not available. High dispersion spectra were obtained 
using the High Resolution Echelle Spectrometer (HIRES) on Keck I and the Magellan Inamori Kyocera Echelle (MIKE) 
on the Magellan Clay telescope. The spectral types of the sample range from B8 to M8 and are nearly equally 
represented by disk-bearing and non-excess sources as determined by {\it Spitzer} 4.5--24 $\mu$m observations.  

This paper is organized as follows: In Section 2 we describe the Upper Scorpius spectroscopic sample and 
discuss the HIRES and MIKE observations and reduction. We then discuss (Section 3) the determination of projected rotational 
velocities and the accretion diagnostics adopted for this analysis. In Section 4 we examine correlations between $v$sin$i$ and 
the presence of circumstellar disks and accretion. We also compare the Upper Scorpius sample with a survey of Taurus-Auriga 
members over a similar range of spectral types. We discuss the implications of our findings as they relate to the disk-braking
scenario and in Section 5 provide a summary of our conclusions.

\section{Observations}

\subsection{Upper Scorpius Membership Sample}
The Upper Scorpius sample was drawn from the {\it Spitzer} 4.5--16 $\mu$m photometric survey of 
Carpenter et al. (2006) and the optical imaging and spectroscopic survey of Slesnick et al. (2006; 2008).
The Carpenter et al. (2006) membership list was compiled from {\it Hipparcos} astrometry for the 
early-type (B--A) stars (de Zeeuw et al. 1999), color-magnitude diagrams with follow-up \ion{Li}{1} 
$\lambda$6708 observations for G--M type stars (Preibisch \& Zinnecker 1999; Preibisch et al. 2002), 
and X-ray detected G--M type stars with \ion{Li}{1} $\lambda$6708 confirmation (Walter et al. 1994; 
Mart\'{i}n 1998; Preibisch et al. 1998; Kunkel 1999; K\"{o}hler et al. 2000). Given that the membership 
selection criteria were based upon stellar properties (proper motions, X-ray activity, \ion{Li}{1} 
$\lambda$6708 absorption) unrelated to circumstellar disks, it is believed that the sample is unbiased 
toward the presence or absence of disks. 

The very low-mass stars and brown dwarfs included in the spectroscopic sample were drawn from the deep, 
wide-field $R-$ and $I-$band imaging survey of Slesnick et al. (2006; 2008), who identified 43 sources that 
have surface gravity signatures consistent with association membership, including 30 brown dwarfs. 
{\it Spitzer} 4.5--24 $\mu$m photometry for many of these sources are presented in Slesnick (2007).
This sample is also considered to be unbiased toward the presence or absence of circumstellar disks.

The final spectroscopic sample is composed of 43 disk-bearing stars identified by Carpenter et al. (2006; 2009)
and 37 non-excess members selected on the basis of having closely matched 
spectral types with the disk-bearing sample. Also included are 3 brown dwarfs (spectral types $\ge$M5.5)
with mid-infrared excess and 4 comparable non-excess sources from Slesnick (2007). Finally 7 late-type
stars and brown dwarfs that lack mid-infrared observations, but which were found to have strong H$\alpha$
emission in low resolution spectroscopic observations (Slesnick et al. 2008) were included. Here we
confirm with high resolution spectroscopy that 6 of these objects exhibit H$\alpha$ emission consistent
with accretion (see \S 3.4). By spectral type, the sample is composed of 20 early-type (B8--A9) stars 
(10 debris disk systems and 10 non-excess sources), 24 F2--K9 type stars (6 primordial disk systems, 5 debris 
disk systems, and 13 non-excess stars); and 50 M-type sources (21 primordial disk systems, 4 debris disk systems, 
18 non-excess sources, and 7 without {\it Spitzer} observations). Provided in Table 1 are details of the spectroscopic sample, organized by spectral 
type taken from Carpenter et al. (2006) and Slesnick et al. (2006; 2008). Also provided is the {\it Spitzer} [4.5], 
[8.0], and [24.0] photometry of Carpenter et al. (2006), Carpenter et al. (2009), and Slesnick (2007) 
as well as the disk classification (i.e. primordial, debris, non-excess) for each source.

\subsection{High Resolution Echelle Spectrometer Observations}
The High Resolution Echelle Spectrometer (HIRES) is a grating cross-dispersed spectrograph permanently 
mounted on the Nasmyth platform of Keck I (Vogt et al. 1994). High dispersion spectra were obtained for
50 Upper Scorpius members on the nights of 2006 June 16, 2007 May 24--25, 2011 March 19, and 2011 April 24. The nights 
were photometric with seeing conditions varying from 0.6--0.9\arcsec. HIRES was configured with the red 
cross-disperser and collimator in beam for increased sensitivity at longer wavelengths. The B2 and C5 
deckers (0.574\arcsec$\times$7.0\arcsec\ and 1.148\arcsec$\times$7.0\arcsec), which have projected slit 
widths of 2 and 4 pixels, were used to provide spectral resolutions of $\sim$66,000 and 37,500, respectively. 
Near complete spectral coverage from $\sim$4800--9200 \AA\ was achieved, a region that includes many 
gravity and temperature sensitive photospheric features as well as permitted and forbidden transitions 
generally associated with accretion or chromospheric activity: H$\beta$, \ion{He}{1} $\lambda$5876, 
\ion{O}{1} $\lambda$6300, H$\alpha$, [S II] $\lambda\lambda$6717, 6730, and \ion{Ca}{2} $\lambda\lambda$8498, 
8542, and 8662. The 3-chip mosaic of MIT-LL CCDs with 15 $\mu$m pixels was used in low gain mode resulting 
in readout noise levels of 2.8, 3.1, and 3.1 e$^{-1}$, for the red, green, and blue detectors, respectively. 
Internal quartz lamps were used for flat fielding and ThAr lamp spectra were obtained for wavelength 
calibration. Given the range of apparent magnitudes for the observed sample, integration times varied 
from 60 to 2400 s with typical signal-to-noise ratios of $\sim$30, 50, and 100 being achieved on the blue, 
green, and red chips, respectively. Radial velocity and spectral standards were observed for a range of 
spectral types. The HIRES observations were reduced using the MAuna Kea Echelle Extraction, {\it makee}, 
reduction script written by Tom Barlow. Makee is publically available from a link provided on the HIRES
webpage.

\subsection{Magellan Inamori Kyocera Echelle Observations}
The Magellan Inamori Kyocera Echelle (MIKE: Bernstein et al. 2003) is a double echelle spectrograph on 
the Magellan Clay telescope at Las Campanas Observatory. High dispersion red and blue spectra were obtained 
for 44 stars and brown dwarfs in Upper Scorpius using MIKE over 4 nights from 
2008 May 12--15. The nights were photometric and seeing conditions varied from $\sim$0.5--1.0\arcsec. MIKE 
achieves near-complete spectral coverage from $\sim$3500--9400 \AA\ in two simultaneous observations through 
red ($\lambda\sim$5000--9400\AA) and blue ($\lambda\sim$3500--5000\AA) cameras. The slit width and binning 
were varied depending upon the brightness of the source, but the majority of the Upper Scorpius members were 
observed using the 0.7\arcsec\ slit and 2$\times$2 binning, yielding a nominal spectral resolution of 
$\sim$30,000. The faintest brown dwarf sources were observed with the 1.0\arcsec\ slit for a spectral
resolution of $\sim$22,000. Integration times for the observed sample ranged from 30 to 2400 s. Internal 
quartz lamp and ThAr spectra were obtained throughout the nights for flat-fielding and wavelength calibration. 
Quartz flats and milky flats were obtained at the beginning of each night. The spectra were reduced and 
extracted using the MIKE pipeline, written by Dan Kelson with methods described in Kelson (2003). The 
MIKE pipeline is a stand-alone Python script specifically written to pre-proceses and extract MIKE data. 

\section{Analysis and Results}
\subsection{Projected Rotational Velocities}
To derive projected rotational velocities, the HIRES and MIKE echelle spectra were first
examined to identify orders and spectral regions that are free of telluric features such
as the O$_{2}$ bands near $\lambda\lambda$6280-6310\AA\ , $\lambda\lambda$6860-6960\AA\ , 
and $\lambda\lambda$7580-7700\AA\ , or the water absorption bands between 
$\lambda\lambda$7150-7300\AA. Echelle orders containing strong emission features related to 
accretion or chromospheric activity (e.g. H$\beta$, H$\alpha$, [O I] $\lambda$6300) were 
also excluded. The remaining orders were then vetted to ensure that high signal-to-noise 
levels were achieved and that an abundance of photospheric absorption features were present. 
The echelle orders selected from the HIRES observations included, but were not limited to 
those spanning spectral ranges from: $\lambda\lambda$5620-5710, $\lambda\lambda$5700-5800,
$\lambda\lambda$5900--6100, $\lambda\lambda$6100--6140, 
$\lambda\lambda$6320--6420, and $\lambda\lambda$6430--6540. The echelle orders selected from the MIKE spectra
included those spanning spectral ranges from: $\lambda\lambda$5640--5805, $\lambda\lambda$5940--6111,
$\lambda\lambda$6045--6220, $\lambda\lambda$6155--6335, and $\lambda\lambda$6270--6450.
Bluer orders were included for more luminous sources.
The minimum measurable $v$sin$i$ scales from the velocity resolution of the observations.
Given that line widths essentially add in quadrature, we conservatively estimate the minimum 
measurable $v$sin$i$ to be $\sqrt{2}$ times lower than the velocity resolution of the
observation or $\sim$3.2 and 5.7 km s$^{-1}$ for the two slit widths used for the
HIRES spectra and $\sim$7.1 and 9.6 km s$^{-1}$ for those adopted for the MIKE observations.

The projected rotational velocities for the late-type (F--M) Upper Scorpius members were 
determined by fitting the target spectra with artificially broadened template spectra of 
slowly rotating ($\le$ 3.3 km s$^{-1}$) standard stars of similar spectral type. The sources 
adopted as templates include: HD100180 (G0) $v$sin$i$=3.3 km s$^{-1}$ 
(Valenti \& Fischer 2005), HD 92788 (G5) $v$sin$i$=0.3 km s$^{-1}$ (Valenti \& Fischer 2005), 
HD 114386 (K3) $v$sin$i$=0.6 km s$^{-1}$ (Valenti \& Fischer 2005), 
GJ 424 (M0) $v$sin$i$ $< 2.5$ km s$^{-1}$ (Browning et al. 2010), Gl 555 (M4) $v$sin$i$ $< 2.5$ km s$^{-1}$ 
(Browning et al. 2010), Gl 876 (M4) $v$sin$i$$<2.0$ km s$^{-1}$ (Delfosse et al. 1998), and 
Gl 406 (M6.5) $v$sin$i$ $< 2.5$ km s$^{-1}$ (Browning et al. 2010). These standards are main 
sequence stars and as such have higher surface gravities (log$g\sim5$) than the typical pre-main sequence
object (log$g\sim4$). The lower densities present in the extended atmospheres of T Tauri stars will result 
in narrower photospheric absorption features, but the effects of rotational broadening should dominate 
over both natural and pressure broadening.

The templates were artificially broadened over a range of $v$sin$i$ values from 2 to 72 km s$^{-1}$
in steps of 2 km s$^{-1}$. For rapid rotators, the upper limit was increased to 400 km s$^{-1}$.
For each artificially broadened spectrum, 
a correlation function was determined between the object and the broadened template. A calibration 
function was then produced coupling the correlation function widths to the $v$sin$i$ values of the 
broadened templates. Using the calibration function and the known peak width of the original
correlation function (i.e. the correlation function between the object spectrum and the unbroadened
template), the object's projected rotational velocity was determined for each echelle order. We 
show in Figure 1 (top panel) an example of the correlation function of an object spectrum and 
template with the rotationally broadened correlation functions overplotted; (middle panel) the
empirical $v$sin$i$ as a function of peak width relation; and (bottom panel) the best-fitting 
rotationally broadened template and observed spectra. The mean $v$sin$i$ over the optimal echelle orders
was adopted as the $v$sin$i$ of the target. The standard deviation of the $v$sin$i$ values over the 
optimal echelle orders serves as the uncertainty of the measurement.

For the early-type (B8--A9) Upper Scorpius members, $v$sin$i$ values were determined using one or 
two echelle orders given the few photospheric absorption features attributable to metals present
in their spectra. For the MIKE observations, the orders containing \ion{Mg}{2} $\lambda$4481 \AA\ and
\ion{Si}{2} $\lambda\lambda$6347, 6471 \AA\ were used in the cross-correlation analysis. Due to the
presence of a blue blocking filter, the HIRES observations were limited only to the \ion{Si}{2}
$\lambda\lambda$6347, 6471 \AA\ features. To serve as template spectra, solar metallicity 
Kurucz (1979) models having similar effective temperatures ($T_{eff}$) as the Upper Scorpius 
sources were generated using the stellar spectral synthesis program SPECTRUM written by R. O. Gray.
These models adopt main sequence surface gravities 
(log$g=5$), microturbulence velocities of 2 km s$^{-1}$, limb darkening coefficients of 0.6, and 
were created with the spectral resolution of the HIRES and MIKE observations. 
As with the late-type stars, a calibration function was used to relate the correlation function 
widths to the $v$sin$i$ values of the early-type stars. Uncertainties for these $v$sin$i$ values
were estimated using models of slightly higher and lower ($\pm$500K) effective temperature. The
dispersion present in the resulting $v$sin$i$ values was then adopted as the uncertainty of the measurement.

The derived $v$sin$i$ values and their associated errors are provided in 
Table 1. For comparison, we present in Table 2 $v$sin$i$ values taken from the literature (e.g. 
Slettebak 1968; Walter et al. 1994; Sartori et al. 2003; Torres et al. 2006; Rice et al. 2010)
for $\sim$15 common sources. In general there is relative agreement with published velocities
with some exceptions, particularly among the rapidly rotating early-type stars.

\subsection{Radial Velocities and Velocity Dispersion}
The relative radial velocity shift between the object and template spectra was determined by the 
cross-correlation analysis (e.g. the top panel of Figure 1). The heliocentric radial velocities 
of the standard stars were taken from the literature (e.g. Valenti \& Fischer 2005; Browning et al. 2010; Nidever et al. 2002; 
Delfosse et al. 1998). Shown in Figure 2 are the distributions of radial velocities for all 
stars and brown dwarfs included in the spectroscopic sample, binned by spectral type. The median 
and median absolute deviation of the entire distribution is $-$6.28$\pm$1.72 km s$^{-1}$, consistent
with the radial velocity reported by de Zeeuw et al. (1999) for the Upper Scorpius OB association,
$-$4.6 km s$^{-1}$. There are, however, several outliers, particularly among the early type stars
that may be single-line, spectroscopic binaries or possibly non-members. The median radial velocity
and the 1$\sigma$ velocity dispersion for each spectral type bin are: B8--A9: $-$6.14$\pm$15.67 km s$^{-1}$,
F2--K9: $-$6.31$\pm$4.61 km s$^{-1}$, and M0--M8: $-$6.28$\pm$3.04 km s$^{-1}$. These values are 
consistent with each other and with the accepted radial velocity of the association from de Zeeuw et al. (1999). In Table 2 
we provide measured radial velocities for $\sim$17 sources taken from the literature. The largest
disparities are found among the early-type stars, several of which are known spectroscopic binaries 
compiled by Kouwenhoven et al. (2007) and one, HIP 78207, is a confirmed double-line spectroscopic 
binary (\S3.3).

\subsection{New and Candidate Spectroscopic Binaries}
The high dispersion spectra revealed 1 early-type and 3 late-type double-line spectroscopic 
binaries that were previously unknown. The classical Be star and possible debris disk system HIP 78207 
exhibits 2 sets of absorption features separated by $\sim$40 km s$^{-1}$. The assumed primary, 
distinguished by broad \ion{He}{1} absorption lines, appears to be rapidly rotating and has a 
radial velocity more consistent with that of the Upper Scorpius OB association. The secondary
component exhibits numerous absorption features attributable to \ion{Fe}{1}, \ion{Fe}{2}, and 
\ion{Si}{2} and has a measured $v$sin$i$ of $\sim$38 km s$^{-1}$. Both H$\alpha$ and H$\beta$ 
exhibit broad emission profiles, consistent with the star's Be classification. The late-type
double-line spectroscopic binaries include HIP 79462 (G2), J160341.8$-$200557 (M2), and 
J160545.4$-$202308 (M2). HIP 79462 is a debris disk system that appears to be a near-equal
mass binary. The velocity separation of the two components is $\sim$63 km s$^{-1}$ with a 
systemic velocity of $-$4.99 km s$^{-1}$ (assuming equal mass). The non-excess source J160341.8$-$200557 also
appears to be a near-equal mass binary with a velocity separation of $\sim$66 km s$^{-1}$
and a systemic velocity of $-$3.71 km s$^{-1}$ (assuming equal mass). The primary and secondary of the primordial 
disk source J160545.4$-$202308 are separated by $\sim$40 km s$^{-1}$, with one component
exhibiting narrow lines and the other somewhat broader.

Given the single observation per source, confirmation of single-line spectroscopic binaries is 
not possible. Spectroscopic binary candidates, however, can be identified by comparing the 
star's radial velocity to that of the association mean; radial velocity dispersions within 
star forming regions are typically $\sim$2 km s$^{-1}$ (e.g. Hartmann et al. 1986).
Five candidate binaries are noted as having broadened cross-correlation functions or 
by absorption line profiles that appear to exhibit two minima. The correlation functions 
of these candidates are not significantly asymmetric, which would suggest that they are 
spectroscopically resolved. The possible spectroscopic binaries include: HD 142987 (G4),
J160823.2$-$193001 (K9), J155918.4$-$221042 (M4), SCH162007.6$-$235915.2 (M6), and
SCH1622351.58$-$231727 (M8). Five other candidate binaries are identified 
on the basis of radial velocities that are at least $\sim$3$\sigma$ from the mean of the 
of the association ($-$6.28 km s$^{-1}$). These candidates include: HIP 79739 (reported as
a binary by Kouwenhoven et al. 2007), HIP 79124 (also reported as a binary by Kouwenhoven 
et al. 2007), HIP 78963, HIP 80130, and HIP 82319.

Several visual binaries are also noted in Table 1. Some of these have been resolved by high angular
resolution imaging techniques: $[$PZ99$]$J161411.0$-$230536 (Metchev \& Hillenbrand 2009),
ScoPMS 31 (K\"{o}hler et al. 2000), and J160643.8-190805 (Dahm, in preparation). These
stars are spatially unresolved in the HIRES spectra and their measured velocities are
potentially biased by light from two or more sources. Other stars were noted as visual binaries on the HIRES guide camera
images (e.g. J160545.4$-$202308, J160702.1$-$201938, J160611.9$-$193532). Follow-up observations
are needed to confirm these as physically associated pairs.

\subsection{Accretion Diagnostics}
In the magnetospheric accretion model, gas from the inner disk rim is channeled along magnetic
field lines to an impact point on the stellar photosphere (Valenti et al. 1993; Hartmann et al. 1998; 
Muzerolle et al. 1998). The infalling gas is inferred from inverse P Cygni line profiles and broadened 
emission lines of \ion{H}{1}, \ion{He}{1}, and \ion{Ca}{2}, often exceeding several hundred km s$^{-1}$ 
(Hamann \& Persson 1992; Batalha \& Basri 1993).

In our Upper Scorpius spectroscopic sample only one early-type star exhibits H$\alpha$ emission, 
the spectroscopic binary HIP 78207 (B8). This source was identified as a classical Be star by Hernandez et al. (2005) 
based upon its lack of near-infrared excess and the presence of strong H$\alpha$ emission. Emission in 
classical Be stars is thought to arise from a gaseous disk that is unrelated to accretion. The traditional
interpretation of the Be star phenomenon is that they are slightly evolved and undergoing mass loss. The
mid-infrared excess associated with this source may arise from hydrogen emission and not from a debris
disk.

The remaining Upper Scorpius sources that exhibit H$\alpha$ emission are late-type stars and brown 
dwarfs (G--M spectral types). Traditionally the distinction between CTTS and WTTS was based upon the 
equivalent width of H$\alpha$, with the demarcating W(H$\alpha$) value established at 10 \AA (for a review see Appenzeller \& Mundt 1989).
Clear differences in the processes responsible for emission are recognized for CTTS (accretion)
and WTTS (chromospheric activity).
Because of the contrast effect (Basri \& Marcy 1995; White \& Basri 2003), W(H$\alpha$) is spectral 
type dependent. Consequently, no unique W(H$\alpha$) value is capable of distinguishing all accreting
sources from non-accreting objects. Various spectral type dependent W(H$\alpha$) criteria have been 
proposed (e.g. Mart\'in 1998; White \& Basri 2003), but the full width of H$\alpha$ emission line 
profiles at 10\% of peak flux was demonstrated by White \& Basri (2003) to effectively distinguish 
between optically-veiled and non-veiled pre-main sequence stars. Full widths of $>$270 km s$^{-1}$ 
were found to imply accretion, independent of spectral type. Jayawardhana et al. (2003) adopt a less 
conservative 10\% H$\alpha$ full width of $>$200 km s$^{-1}$ to indicate accretion among very low mass 
stars and brown dwarfs (M5--M8 spectral types) in IC\,348 and Taurus-Auriga. Natta et al. (2004) conclude 
that H$\alpha$ 10\% full widths of 200 km s$^{-1}$ for such sources correspond to mass accretion rates 
($\dot{M}$) of $\sim$10$^{-11}$ M$_{\odot}$ yr$^{-1}$.

We adopt a composite criterion for accretion, requiring full widths of $>$270 km s$^{-1}$ for spectral 
types earlier than M5 and $>$200 km s$^{-1}$ for types M5 and later. To determine whether the sources 
in Upper Scorpius are accreting, we first estimate the continuum level by linearly interpolating between 
regions on either side of H$\alpha$ that are free of photospheric absorption features and unaffected by 
the broadened wings of strong H$\alpha$ emission profiles. The peak level of H$\alpha$ emission is then 
determined relative to the defined continuum level. The limits of the 10\% width of peak emission were 
then defined and the corresponding velocities calculated, which are provided in Table 1. 

Just 18 of the 74 late-type (F2--M8) stars and brown dwarfs in the Upper Scorpius sample meet the 
modified accretion criterion, including two marginal sources: the K2-type star [PZ99]J160421.7$-$213028 
(H$\alpha$ velocity width $\sim$252 km s$^{-1}$, but exhibits an inverse P Cygni profile characteristic 
of accretion) and the M4-type star J155729.9$-$225843 ($\sim$238 km s$^{-1}$). The G4-type, non-excess
star HD 142987 exhibits broad H$\alpha$ emission (H$\alpha$ velocity width $\sim$600 km s$^{-1}$), but 
is a rapid rotator and suspected spectroscopic binary. Three additional sources that are M5 or later
(J160611.9$-$193532, SCH16200756$-$23591522, and SCH16235158$-$23172740) exhibit H$\alpha$ velocity 
widths $>$200 km s$^{-1}$, but are also rapid rotators. Shown in Figure 3 are the emission line profiles
for each of the suspected accretors, arranged by spectral type. By spectral type, 0 of 4 F-type stars,
0 of 6 G-type stars, 3 of 14 K-type stars, 11 of 40 M0--M5 stars, and 4 of 10 M5.5--M8 sources have
H$\alpha$ velocity widths that are consistent with accretion and not the result of rotational broadening. 
The steady increase in accretion fraction with spectral type or mass from zero for F and G-type stars 
to $\sim$40\% for the latest spectral types parallels the increase in primordial disk fraction.

\section{Correlations with Rotation}

\subsection{Rotation and Stellar Mass}
The dependence of rotation upon stellar mass is a well-established correlation among normal main 
sequence stars such that rotational velocities peak near $\sim$B5 spectral types before declining 
sharply near mid-F (e.g. Slettebak 1955; Kraft 1967; Tassoul 2000). This decline is attributed to
the development of a convection zone, from which a magnetic field can be a generated and 
coupled with a stellar wind. The wind interaction effectively brakes the star, causing a decline 
in rotation rate with main sequence age ($\tau$) such that $v$sin$i$ $\propto$ $\tau^{-1/2}$ 
(Schatzman 1962; Skumanich 1972). The efficiency of angular momentum loss induced by magnetic activity 
is mass dependent (Scholz et al. 2007). This is observationally supported by the steady progression 
of remnant rapid rotators toward later spectral types with age: G--M in the $\sim$50 Myr $\alpha$ Persei 
cluster, mid-K through M in the $\sim$100 Myr old Pleiades, and M-type only in the $\sim$600 Myr Hyades 
(Delfosse et al. 1998 and references therein).

To examine the mass-dependence of rotation at $\sim$5 Myr, we show in Figure 4 the projected rotational 
velocities of the Upper Scorpius sample as a function of spectral type, a proxy of stellar mass. At this 
age, a solar mass star would correspond to a spectral type of $\sim$K7 using the pre-main 
sequence models of Baraffe et al. (1998) or Siess et al. (2000). A general decline in rotational velocity 
is apparent from B8 to mid-K spectral types, but appears to be much shallower than is evident among main
sequence stars. Figure 4 suggests that the average $v$sin$i$ declines from $\sim$100 km s$^{-1}$
to $\le$20 km s$^{-1}$ between $\sim$F2 and K5 and remains at this level into the substellar mass regime. 

Among early-type (B8--A9) stars, $v$sin$i$ spans a broad range of values from $<$50 to $\sim$400 km s$^{-1}$. 
This is comparable to the distribution of $v$sin$i$ values for B0--B9 dwarfs in the $\sim$13 Myr old 
galactic clusters h \& $\chi$ Persei (Strom et al. 2005). The $v$sin$i$ values for late-type stars in 
Upper Scorpius are similar to those in Taurus-Auriga and Chamaeleon I ($\sim$11--24 km s$^{-1}$, 
Nguyen et al. 2009), $\eta$ Cha ($\sim$5--20 km s$^{-1}$), TW Hydra ($\sim$5--32 km s$^{-1}$), and 
the $\beta$ Pictoris moving group ($\sim$5--24 km s$^{-1}$, Scholz et al. 2007). We note the presence
of a small number of rapid rotators ($v$sin$i$$\ge$50 km s$^{-1}$) among K and M-type Upper Scorpius
members, including one ultrafast ($v$sin$i$$\sim$100 km s$^{-1}$) rotator, [PZ99]J160042.8$-$212737 (K7).

Shown in Figure 5 are boxplots of $v$sin$i$ for the Upper Scorpius sample grouped into bins for 
B8--A9, F2--K4, K5--K9, M0--M5, and M5.5--M8 spectral types. The F2--K4 and K5--K9 bins straddle
the stellar mass boundary where the sharp decline in rotational velocity occurs among main sequence stars.
Spectral types earlier than $\sim$K4, corresponding to masses of $\sim$1.4 M$_{\odot}$, will not have 
convective zones when on the main sequence as early F-type stars.

The median $v$sin$i$ are: 195 ($\pm$70), 37.8 ($\pm$7.4), 13.8 (+21.3,$-$8.2), 16.52 ($\pm$5.3),
and 17.72 ($\pm$8.1) km s$^{-1}$ for the B8--A9, F2--K4, K5--K9, M0--M5, and M5.5--M8 spectral type
bins, respectively. The uncertainties presented are the median absolute deviations. The rotational 
velocities appear to bottom out between K5 and K9 before rising slightly toward later spectral types.
The distribution of rotational velocities for the brown dwarfs is notably distinct from that of 
M0--M5 type stars, showing a substantial dispersion, possibly the result of unresolved spectroscopic
binaries among the rapid rotators. These M5 and later spectral types will evolve into late-M and
L-type dwarfs, which in the field are more rapidly rotating than early to mid-M type dwarfs
(Reiners \& Basri 2010; Blake et al. 2010). If at this age such sources are already spinning
more rapidly, the difference may not be attributable to a Skumanich type spin-down mechanism.

Nguyen et al. (2009) find clear differences in the $v$sin$i$ distributions of F--K stars (24--26 km s$^{-1}$) 
and M dwarfs (11 km s$^{-1}$) in Taurus-Auriga and Chamaeleon I that are statistically significant. 
Employing the non-parametric Kolmogorov-Smirnov (KS) test to the $v$sin$i$ values of our samples of 
early-type (B8--A9) and late-type (F2--M8) stars, we find a small (10$^{-5}$) probability (as defined
by Press et al. 1986) that the two samples are drawn from the same parent population.  Between the
F2--K9 type stars and M-dwarfs, however, the $v$sin$i$ distributions are only marginally  distinguishable
with a $\sim$10\% probability that the two samples are drawn from the same parent population. One possible
explanation for this difference with the results of Nguyen et al. (2009) is that the Taurus and Chamaeleon I samples are
strongly weighted toward spectral types earlier than M4. Only 6 of their 144 sources have types later 
than M4 compared with 28 of 94 sources in the Upper Scorpius sample.

Given the long timescales associated with the $v$sin$i$ $\propto$ $\tau^{-1/2}$ stellar wind braking 
mechanism of Skumanich (1972), another process must be responsible for the steady decline of rotation 
rate with mass observed in pre-main sequence stars. Two prominent explanations include 1) star-disk 
interactions that regulate stellar rotation and 2) changes in internal stellar structure that enhance 
magnetic activity (Scholz et al. 2007). Growing evidence suggests that disk dissipation timescales are 
strongly mass-dependent (e.g. Lada et al. 2006; Carpenter et al. 2006; Dahm \& Hillenbrand 2007; 
Slesnick 2007), such that late-type stars (K+M) are capable of retaining their primordial disks for 
prolonged periods relative to their more massive counterparts. 
Scholz et al. (2007) suggest that at ages $>$5 Myr, K-type stars have deeper 
convection zones than F--G type stars, implying that rotational braking becomes more effective for 
these later spectral types. The M-dwarfs, however, remain fully convective, suggesting that the physical 
mechanism responsible for generating magnetic fields in these stars must be distinct from an $\alpha$-$\omega$ 
type dynamo. 

\subsection{Rotation and Infrared Excess}
The correlation between rotation and circumstellar disks in the $\sim$1 Myr old ONC found
by Rebull et al. (2006) predicts that disk-bearing and non-excess stars in more evolved clusters 
should exhibit significantly different period distributions. Hartmann (2002) 
suggests that the timescales of disk clearing and subsequent spin-up may be comparable. If stars 
have recently cleared their inner disks, they may still be rotating slowly as though they were 
experiencing disk-braking. Rebull et al. (2006) predict that a lower fraction of slowly rotating 
stars lacking disks should be present in older clusters given that stars released from disk locking 
should have had sufficient time to spin-up. 

To search for possible correlations between rotation and infrared excess in Upper Scorpius, we
merge the projected rotational velocities presented in \S3.1 with the {\it Spitzer} IRAC and MIPS 
photometry of Carpenter et al. (2006; 2009) and Slesnick (2007). The {\it Spitzer} survey of Upper Scorpius includes
only IRAC channels 2 and 4, limiting our mid-infrared color selection to [4.5]$-$[8.0] (Rebull et al. 
2006 present [3.6]$-$[8.0] colors in the ONC IRAC survey). In addition to the {\it Spitzer} 
photometry of the Upper Scorpius members, we also adopt the disk classifications of
Carpenter et al. (2006; 2009) and Slesnick (2007). Carpenter et al. (2009) assign a primordial disk
classification to K+M type stars that exhibit 8 and 16 $\mu$m excesses. Sources with weak excesses
at 16 or 24 $\mu$m were classified as debris disk sources. Slesnick (2007) defines brown dwarfs
with [3.6]$-$[8.0] color excess emission $>$3$\sigma$ above photospheric levels as 
disk-bearing (assumed here to be primordial) sources. We include both primordial and debris
disk systems in the disk-bearing sample given that the interiors of debris disks may have only 
recently dissipated. We also exclude the few double-line spectroscopic binaries from the
statistical analysis. In Figures 6a) and 6b) we plot $v$sin$i$ as a
function of [4.5]$-$[8.0] and [4.5]$-$[24.0] colors, respectively, for the spectroscopic sample. 
To minimize mass-dependent rotation 
effects, we divide the sample into 3 spectral type bins: B8--A9, F2--K9, and M0--M8. To 
facilitate comparison with rotational period surveys (e.g. Rebull et al. 2006), we plot along
the right ordinate the rotation periods associated with specific rotational velocities
estimated by assuming a stellar radius typical of the spectral type range: B8--A9 stars
2.5 R$_{\odot}$; F2--K9 stars 2.0 R$_{\odot}$ (from the pre-main sequence models of Baraffe et al. 1998);
and M0--M8 sources 0.88 R$_{\odot}$ (from the same models).

For the early-type stars (B8--A9), we use the KS-test to demonstrate that the disk-bearing and non-excess 
sources are statistically indistinguishable, i.e. the presence or absence of circumstellar disks 
has no bearing on the observed $v$sin$i$ values. These early-type excess sources, however, are 
associated with debris disks having substantial ($\ge$ 10 AU) inner disk gaps (Carpenter et al. 2006, 
Dahm \& Carpenter 2009) and may be second-generation debris disks in which dust has been re-generated
through the collision of planetesimals.

Among the late-type sources, the distributions of disk-bearing stars and brown dwarfs appear to be 
shifted toward lower $v$sin$i$ than their non-excess counterparts. Performing the KS-test on 
the $v$sin$i$ distributions of disk-bearing and non-excess F2--K9 type stars, we find differences that 
are significant, but only at the 97.2\% confidence level. The median $v$sin$i$ for the disk-bearing 
and non-excess F2--K9 type stars are 15.4 and 37.8 km s$^{-1}$, respectively. For the M-dwarfs, however, 
the probability that the $v$sin$i$ distributions of the disk-bearing and non-excess sources are drawn from 
the same parent population is extremely low, $\sim$0.03\%, implying a statistically significant difference. 
This argues in favor of a correlation between rotation and circumstellar disks such that disk-bearing 
sources are rotating more slowly than their non-excess counterparts. The median $v$sin$i$ for 
the disk-bearing and non-excess M-type stars and brown dwarfs are 13.1 and 26.2 km s$^{-1}$, respectively. 

While conflicting with the findings of Nguyen et al. (2009) in Taurus-Auriga and Chamaeleon I, this result
agrees with those of Rebull et al. (2006) for Orion. The discrepancy with Nguyen et al. (2009)
may arise from the substantially lower mass sample included in Upper Scorpius. Shown in Figure 7 are the cumulative 
distribution functions (CDF) of $v$sin$i$ values for the B8--A9, F2--K9, and M0--M8 stars and 
brown dwarfs with and without mid-infrared excess, determined using the Astronomy Survival (ASURV) 
analysis package of Feigelson et al. (1996). The CDFs of disk-bearing and non-excess sources of F2--K9 
type stars and the M-dwarfs are substantially different. Figure 8 demonstrates this further with 
box and whisker plots of the $v$sin$i$ distributions for disk-bearing and non-excess sources, 
ordered by spectral type. The boxes for the disk-bearing F2--K9 and M0--M8 sources are clearly 
displaced relative to those of the non-excess sources. 

There are no locking mechanisms available for debris disk systems, but if recently released
from their inner disks, such stars could still be experiencing the effects of disk braking.
The few late-type debris disks in the sample are nearly evenly divided among slow rotators
($v$sin$i$$\sim$10 km s$^{-1}$) and moderate rotators ($v$sin$i$$\sim$20 km s$^{-1}$). The
single exception is the F2-type debris disk source HIP 79643 ($v$sin$i$$\sim$80 km s$^{-1}$),
which more closely resembles the early-type (B+A) stars. As a whole, the rotational velocity
distribution of late-type debris disk systems cannot be distinguished from that of primordial
disk-bearing sources, however, the full sample of debris-like systems identified by
Carpenter et al. (2009) should be included before more definitive statements are made.
There also appears to be few slow rotators among the non-excess sources (\S 4.4), which supports
the prediction of Rebull et al. (2006) that a lower fraction of long period, non-excess sources
should be present at more advanced ages. 

\subsection{Rotation and Accretion}

Accretion is generally associated with near- and mid-infrared excess emission originating from
hot dust in the inner disk. For 63 of 67 (94\%) pre-main sequence stars in Taurus-Auriga and
Chamaeleon I, Nguyen et al. (2009) find that the presence or absence of accretion as determined by the
H$\alpha$ 10\% velocity width criterion of Jayawardhana et al. (2003) is strongly correlated 
with {\it Spitzer} 8.0 $\mu$m excess. Shown in Figure 9 are the [4.5]$-$[8.0] colors for the
Upper Scorpius sample plotted as a function of H$\alpha$ equivalent width, W(H$\alpha$), and 
H$\alpha$ 10\% velocity width for G4--M4 type stars (upper panels) and M5--M8 type sources (lower
panels). Several non-excess sources fall between the 200 and 270 km s$^{-1}$ boundaries,
presumably a result of rotational broadening. A substantial fraction (14 of 26) of primordial disk
sources do not appear to be accreting.

We next examine the $v$sin$i$ distributions for correlations between accretion and rotation.
Only stars hosting gas-rich primordial disks with magnetospheric footprints in the inner 
disk should experience rotational braking. Considering the late-type (G4--M8) Upper
Scorpius members with H$\alpha$ in emission, we plot in Figure 10 $v$sin$i$ as a function
of H$\alpha$ 10\% velocity width. There are too few (3) suspected accretors among the G4--K9
type stars to perform proper statistical analysis. Applying the KS-test to the M0--M8 accreting
and non-accreting populations, we find that while the maximum deviation between the two 
cumulative distribution functions is substantial, the significance level remains low,
such that the null hypothesis that accreting and non-accreting sources are drawn from the
same parent population cannot be rejected with any degree of confidence.

Nguyen et al. (2009) also find negative KS-test results between accretion and rotation
in Taurus-Auriga and Chamaeleon I, which appear to stem from the presence of a large number
of rapidly rotating accretors having $v$sin$i>$20 km s$^{-1}$. Such objects are rare in
Upper Scorpius with only one accretor having a comparable rotational velocity.
To explain the lack of correlation between accretion and
mid-infrared excess and between rotation and accretion in the Upper Scorpius sample, we
postulate that some primordial disk-bearing sources may be accreting at levels below the
detection threshold of the H$\alpha$ velocity width criterion.
Dahm \& Carpenter (2009) and Dahm (2010) find that many of the primordial disk-bearing stars in
Upper Scorpius exhibit reduced levels of near- and mid-infrared excess emission and order of magnitude
lower mass-accretion rates compared with comparable Class II sources in Taurus-Auriga. They suggest
that the inner disk radii for some of the Upper Scorpius sources exceed their respective dust
sublimation radii, potential evidence for inner disk evolution. Such transition-like objects
may either be weakly accreting or no longer accreting through the inferred inner disk gap.

\subsection{Implications for Disk Braking}
In the disk braking model, rapidly rotating protostars spin down during the T Tauri phase
over the disk braking timescale ($\tau_{DB}$). Ignoring spin-up torques induced by accreting
material, Hartmann (2002) derives a lower limit for $\tau_{DB}$ given by:

$\tau_{DB} \ge 4.5 \times 10^{6} yr M_{0.5} \dot{M}^{-1}_{-8} f$

\noindent where M$_{0.5}$ is the stellar mass in units of 0.5 M$_{\odot}$, $\dot{M}_{-8}$
is the mass accretion rate in units of 10$^{-8}$ M$_{\odot}$ yr$^{-1}$, and $f$ is the 
ratio of the stellar angular velocity to the breakup velocity. The dependence upon mass
accretion rate and the range of $\dot{M}$ values among CTTS of comparable mass and age
implies an intrinsic spread in disk braking timescales (Hartmann 2002).

For typical values of $\dot{M}$ and mass in Taurus-Auriga, $\tau_{DB}$ is $\sim$5 Myr. To
explain the lack of correlation between accretion or the presence of inner disks and rotational 
velocities in Taurus-Auriga and Chamaeleon I, Nguyen et al. (2009) suggest that insufficient 
time has elapsed for disk braking to take hold in these young systems. This, however, 
conflicts with the findings of Rebull et al. (2006) in the comparably aged Orion region.
Alternatively Nguyen et al. (2009) propose that a substantial age dispersion may
be present in these regions that masks the expected signatures of disk braking in the 
rotational velocity distributions.

The age of Upper Scorpius is approximately equal to the disk braking timescale for a half-solar 
mass star, a nominal $\dot{M}$ value of 1.0$\times 10^{-9}$ M$_{\odot}$ 
yr$^{-1}$, and $f>$0.1. The age dispersion within Upper Scorpius is estimated to be $\le$3 Myr 
(Slesnick et al. 2008), eliminating this potential source of uncertainty. The correlations between
the presence of disks and rotation for M-dwarfs and possibly F--K type stars in Upper Scorpius 
support the disk braking hypothesis. Circumstantial evidence is provided by the small
fraction of non-excess sources that are slow rotators. Shown in the upper panel of Figure 11 
are the rotation periods for 464 sources in Orion from Rebull et al. (2006) plotted as 
a function of [4.5]$-$[8.0] color. Defining stars having [4.5]$-$[8.0] colors between
$-$0.2 and +0.2 mag. as non-excess sources, we find that the period distribution for these
stars spans from $\le$1 to $>$25 days. A substantial fraction (63/140 or 45\%) 
of non-excess sources in Orion have rotation periods of $>$5 days. Assuming the Orion
sample to be comprised predominantly of late-type stars having radii $\sim$1.5 $R_{\odot}$,
such periods correspond to equatorial rotation velocities of $\sim$15 km s$^{-1}$
or less. 

Shown in the lower panel of Figure 11 are histograms of $v$sin$i$ for late-type (F2 and 
later), non-excess sources in Upper Scorpius as well as Taurus-Auriga
and Chamaeleon I (from Nguyen et al. 2009). The fraction of slowly rotating
($v$sin$i\lesssim15$ km s$^{-1}$), non-disk bearing sources in Upper Scorpius is found 
to be 5/30 ($\sim$17\%), while that of Taurus and Chamaeleon I is 23/71 ($\sim$32\%).
These first order comparisons with Orion, Taurus-Auriga, and Chamaeleon I are suggestive
of a gradual decline in the number of slowly rotating, non-excess sources with age.

Other factors, however, should be considered before attributing these results to disk
regulation including: normal accretion-induced variability (e.g. Johns \& Basri 1995),
intermittent periods of spin-up or enhanced disk emission from 
rapid accretion events (see Rebull et al. 2006 and references therein); unrecognized
spectroscopic binaries among the rapid rotators; and observational uncertainties including
inclination angle. The role of environment in disk evolution and the effect of nearby 
early-type stars that can ionize and disperse molecular gas in the envelopes and
disks surrounding low mass populations (e.g. Rebull et al. 2006; O'dell 1998) is poorly understood.
The environs of Upper Scorpius is certainly distinct from that found in Taurus-Auriga (a sparsely
populated region with few early-type stars) or the ONC (a densely packed cluster with
several proximal O-type stars).

\section{Summary and Conclusions}

We have carried out an extensive investigation into the influence of circumstellar disks 
upon rotation for 94 stars and brown dwarfs in the $\sim$5 Myr Upper Scorpius OB association
using {\it Spitzer} IRAC and MIPS 24 $\mu$m photometry and measured projected rotational 
velocities from high dispersion spectroscopy. We find that rotational velocity is
strongly correlated with spectral type, a proxy for mass, such that the median $v$sin$i$
for B8--A9 type stars is: 195$\pm$70 km s$^{-1}$, for F2--K4 types: 37.8$\pm$7.4 km s$^{-1}$,
for K5--K9: 13.8$^{+21.3}_{-8.2}$ km s$^{-1}$, for M0--M5: 16.52$\pm$5.3 km s$^{-1}$, and
for M5.5--M8: 17.72$\pm$8.1 km s$^{-1}$. The distribution of rotational velocities for brown
dwarfs (M5.5 and later) exhibits a substantial dispersion that is not present among M0--M5
type stars. This may in part result from unresolved spectroscopic binaries. Late-M and 
L-type dwarfs in the field are more rapidly rotating than early to mid-M type stars. If
this difference is present at $\sim$5 Myr, it would suggest that it is not due to a 
Skumanich type spin-down mechanism.

We find with a probability of $\ge$0.99 that among M-dwarfs, disk-bearing sources rotate
more slowly on average than their non-disk bearing counterparts. Disk-bearing F--K type stars 
also appear to rotate more slowly than non-excess sources, but only at the $\sim$97\%
confidence level. Given that less than half of the F--K type Upper Scorpius members observed
by {\it Spitzer} have $v$sin$i$ measurements available, we suggest that additional high
dispersion spectra are needed to adequately explore this mass range and to better quantify 
this result.

Considering the late-type (F2--M8) stars and brown dwarfs, we find few slowly rotating, 
non-excess sources in Upper Scorpius, suggesting that most have spun up following disk
dissipation. Comparing the rotational velocities of late-type, non-excess sources in 
Upper Scorpius with those in the presumably younger Taurus-Auriga and Chamaeleon I star
forming regions and Orion, we find that the fraction of slow rotators (i.e. $v$sin$i$$\lesssim$15
km s$^{-1}$) is lower in Upper Scorpius by a factor of $\sim$2 or more. This supports the prediction
of Rebull et al. (2006) that lower fractions of slowly rotating, non-excess stars should 
be present in older, more evolved clusters.

Among the small number of late-type (K5--M5), debris disk sources in the Upper Scorpius 
sample, we find that rotational velocities are nearly evenly divided between slow and 
moderate rotators. Such systems may be representative of second generation debris disks
or the remnants of primordial disks with substantial inner holes. These stars may have 
been recently released from their inner disks and are in the process of spinning up.

The lack of a strong correlation between accretion and rotation among the Upper Scorpius
sources possibly originates from the small number of primordial disk systems that are 
accreting at detectable levels using the H$\alpha$ 10\% width accretion criteria of
White \& Basri (2003) and Jayawardhana et al. (2003). Alternatively, given the evidence
for reduced near- and mid- infrared excess emission among Upper Scorpius primordial 
disks, it is possible that many inner disks have receded beyond their respective dust
sublimation radii and are no longer accreting.
Such transition-like objects, if recently released, may still be experiencing the
effects of disk braking and may not have had sufficient time to spin up.

The high dispersion spectra have revealed 4 new double-line spectroscopic binaries in 
Upper Scorpius as well as 5 possible spectroscopic binaries. Additional high resolution spectra are needed
for confirmation and to determine the orbital properties of these systems.

Future investigations of rotation and disks should include the remaining debris disk candidates 
in Upper Scorpius identified by Carpenter et al. (2009) in order to examine whether differences 
exist between the rotational properties of debris disks and primordial disks. Similar
investigations of other young clusters and associations with {\it Spitzer} observations available
are critically needed over a broad range of ages to confirm the role of circumstellar disks in
the regulation of angular momentum in pre-main sequence stars.

\acknowledgments
This work is based on observations made with the {\it Spitzer} Space Telescope, which is operated by the
Jet Propulsion Laboratory (JPL), California Institute of Technology, under NASA contract 1407. The
Digitized Sky Surveys, which were produced at the Space Telescope Science Institute under 
U.S. Government grant NAG W-2166, were used as were the the SIMBAD database operated at CDS, 
Strasbourg, France, and the Two Micron All Sky Survey (2MASS), a joint project of the University of
Massachusetts and the Infrared Processing and Analysis Center (IPAC)/California Institute of Technology,
funded by NASA and the National Science Foundation. We wish to thank Lynne Hillenbrand and John Carpenter 
for many insightful discussions and suggestions that greatly improved this manuscript.

\clearpage
\begin{figure}
\epsscale{0.65}
\hspace{2cm}  \vspace{2cm}  \includegraphics[width=11cm,angle=0]{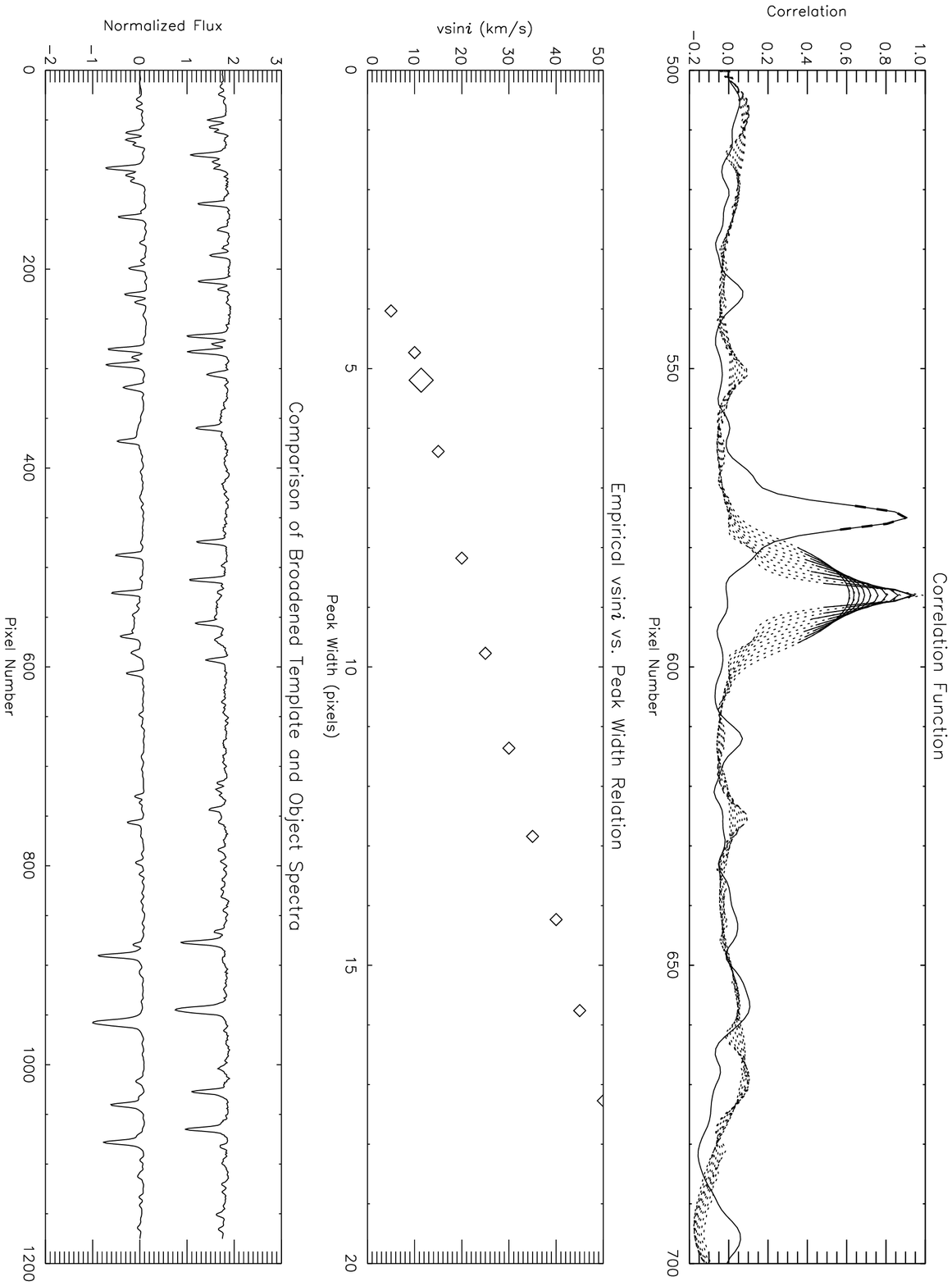}  \vspace{-2cm}    
\caption[f1.ps]{(top panel) The correlation function between an object spectrum and a slowly rotating template spectrum
with the rotationally broadened correlation functions for 10 velocities between 5 km s$^{-1}$ (narrowest profile) and 50 km s$^{-1}$ (broadest profile) superimposed. 
The apparent shift between correlation function and the broadened templates is representative of the radial velocity 
difference between the object and template sources. (middle panel) Empirical $v$sin$i$ as a function of peak width
of the correlation function. The $v$sin$i$ of the best-fitting correlation function is represented by the enlarged
symbol. (bottom panel) The best-fitting, rotationally broadened template spectrum and object spectrum.
\label{f1}}
\end{figure}
\clearpage

\clearpage
\begin{figure}
\epsscale{0.65}
\hspace{2cm}  \vspace{2cm}  \includegraphics[width=11cm,angle=0]{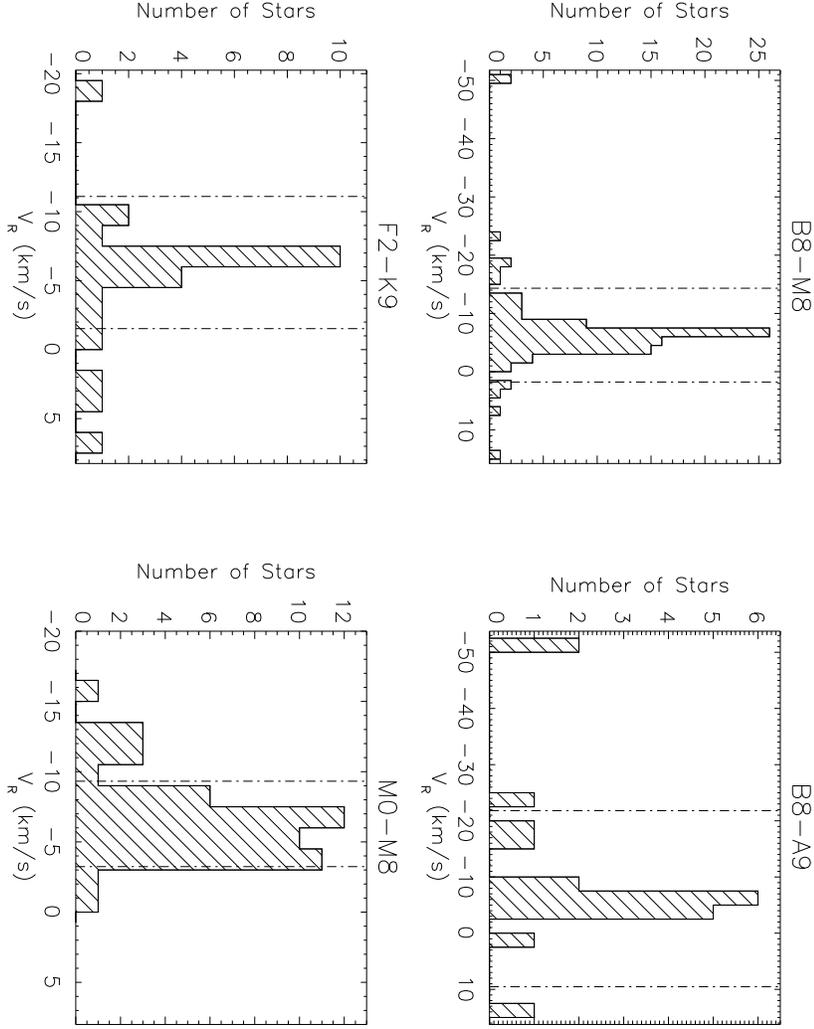}  \vspace{-2cm}    
\caption[f2.ps]{The heliocentric radial velocity distribution for all Upper Scorpius members in the
spectroscopic sample (upper left panel). The median radial velocity is $-$6.15 km s$^{-1}$, consistent with 
results found in the literature for the Upper Scorpius OB Association. In the other panels are the 
heliocentric radial velocity distributions by spectral type: B8--A9 (upper right panel), F2--K9 (lower left panel), and M0--M8 (lower right panel). The 1$\sigma$
dispersions are represented by the vertical dot-dashed lines in each panel. The median
radial velocities for each spectral type bin are consistent with that for all sources. Several outliers, 
however, are present, particularly among the early type stars. These may be spectroscopic binaries or
non-members.
\label{f2}}
\end{figure}
\clearpage

\clearpage
\renewcommand{\thefigure}{\arabic{figure}\alph{subfigure}}
\setcounter{subfigure}{1}
\begin{figure}
\epsscale{0.5}
\hspace{2cm}  \vspace{2cm}  \includegraphics[width=11cm,angle=0]{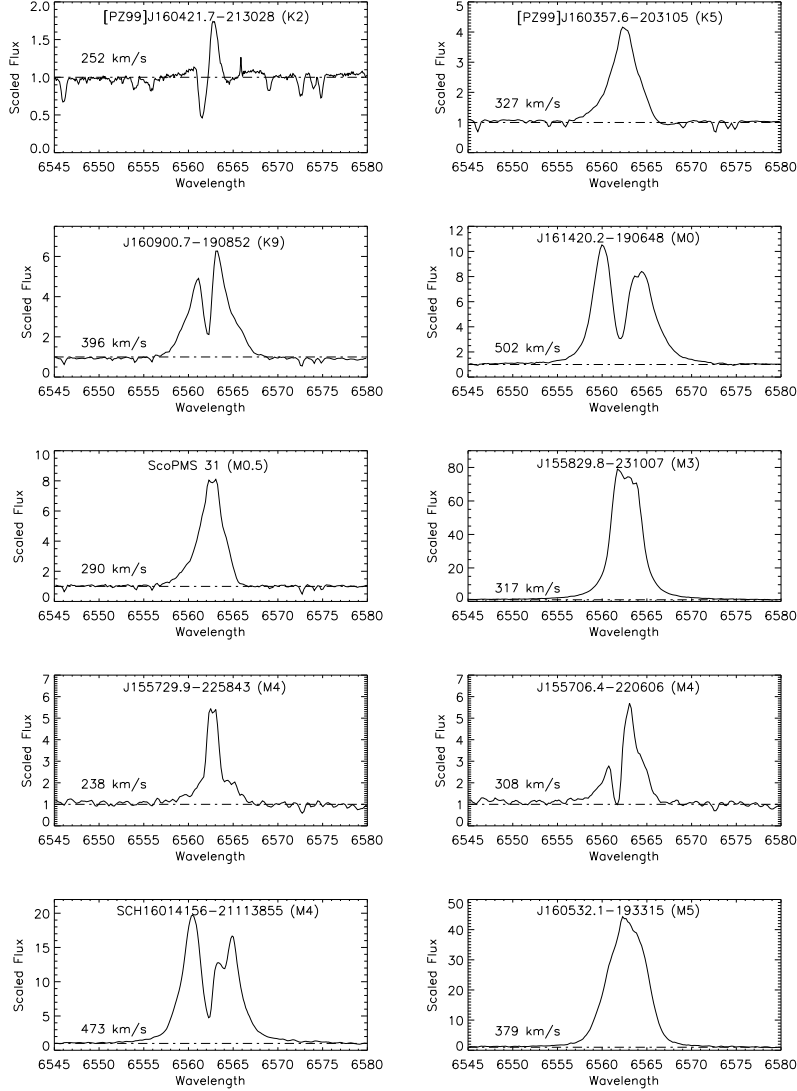} \vspace{-2cm}
\caption[f3a.ps]{H$\alpha$ emission line profiles for each of the suspected accretors
in the Upper Scorpius late-type sample, ordered by spectral type. The assumed continuum
level is represented by the dot-dashed line and the measured velocity widths are given
in each panel. By spectral type, 0 of 4 F-type stars, 0 of 6 G-type stars, 3 of 14 K-type
stars, 11 of 40 M0--M5 stars, and 4 of 10 M5.5--M8 sources have H$\alpha$ velocity widths 
consistent with accretion. Two marginal accreting sources are included here for completeness: 
the K2-type star [PZ99]J160421.7$-$213028 and the M4-type star J155729.9$-$225843.
\label{f3a}}
\end{figure}
\clearpage

\addtocounter{figure}{-1}
\addtocounter{subfigure}{1}
\clearpage
\begin{figure}
\epsscale{0.5}
\hspace{2cm}  \vspace{2cm}  \includegraphics[width=11cm,angle=0]{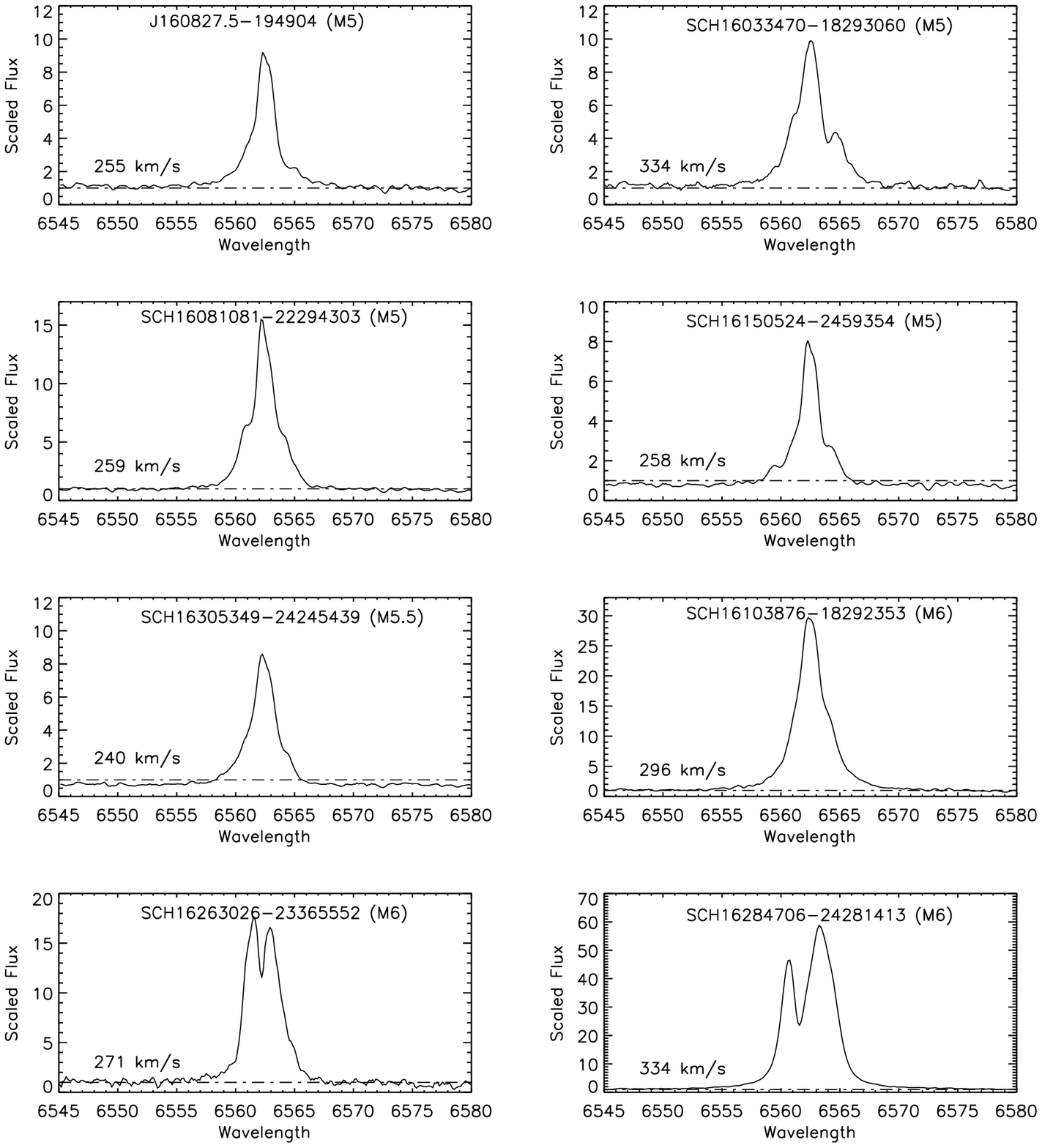}  \vspace{-2cm}    
\caption[f3b.ps]{(continued)
\label{f3b}}
\end{figure}
\clearpage

\clearpage
\begin{figure}
\epsscale{0.65}
\hspace{2cm}  \vspace{2cm}  \includegraphics[width=11cm,angle=0]{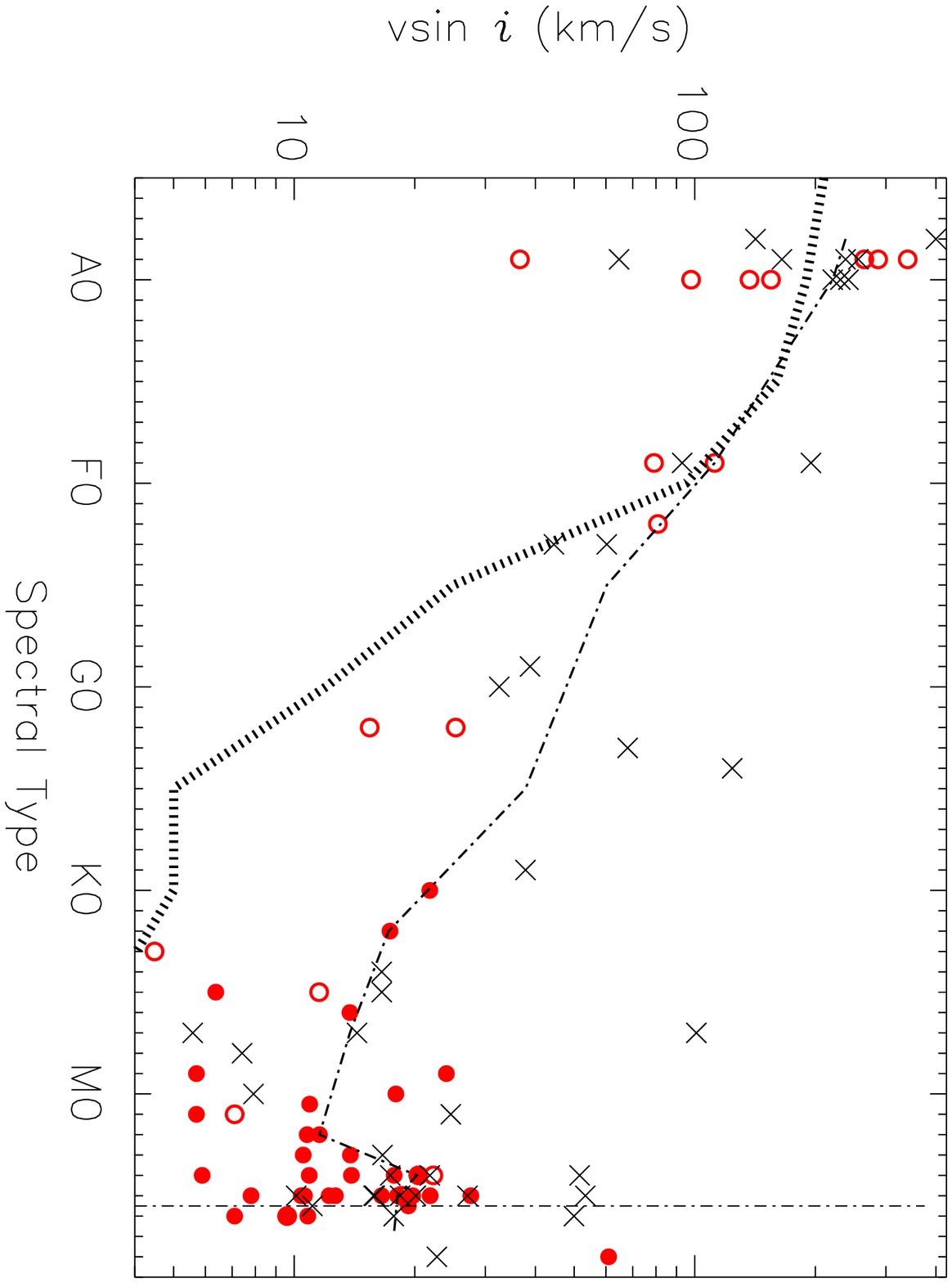}  \vspace{-2cm}    
\caption[f4.ps]{Projected rotational velocity as a function of spectral type for the Upper Scorpius
sample. Primordial disk-bearing sources are represented by solid red circles, debris disk systems as open red
circles, and non-excess sources as crosses. At $\sim$5 Myr, a solar mass star would correspond to a $\sim$K7 spectral
type using the pre-main sequence models of Baraffe et al. (1998). The vertical dot-dashed line represents the
approximate location of the sub-stellar mass boundary, near M5.5 spectral type at the assumed age of Upper Scorpius.
A general decline in rotational velocity is apparent from B8 to early M, paralleling the distribution of main 
sequence rotational velocities, which are represented by the dashed curve. Also shown as a dot-dashed line are the 
median $v$sin$i$ values for various spectral type bins. Near M4, the median rotational velocity appears to rise 
slightly, plateauing near $\sim$18 km s$^{-1}$ well into the substellar mass regime.
\label{f4}}
\end{figure}
\clearpage

\clearpage
\begin{figure}
\epsscale{0.65}
\hspace{2cm}  \vspace{2cm}  \includegraphics[width=11cm,angle=0]{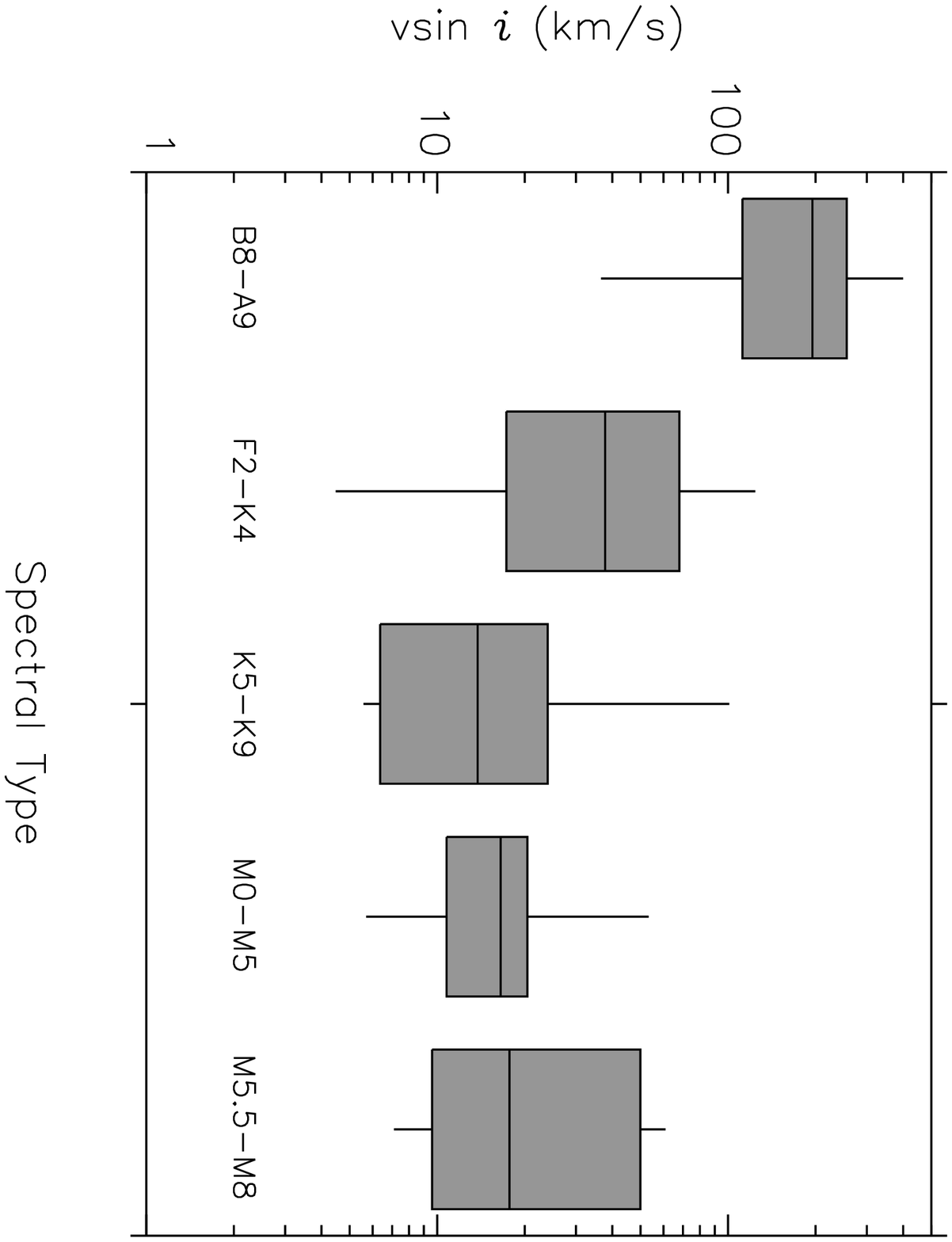}  \vspace{-2cm}    
\caption[f5.ps]{Boxplots of $v$sin$i$ for the Upper Scorpius sample grouped into bins for B8--A9, F2--K4, 
K5--K9, M0--M5, and M5.5--M8 spectral types. The central rectangles represent the first and third quartiles 
of each bin and the ``whiskers'' show the spread between minimum and maximum values. The horizontal lines within
each box represent the median values for the spectral type bins. The median $v$sin$i$ are: 195 ($\pm$70),
37.8 ($\pm$7.4), 13.8 (+21.3,$-$8.2), 16.52 ($\pm$5.3), and 17.72 ($\pm$8.1) km s$^{-1}$ for the B8--A9, 
F2--K4, K5--K9, M0--M5, and M5.5--M8 spectral type bins, respectively. 
\label{f5}}
\end{figure}
\clearpage

\clearpage
\renewcommand{\thefigure}{\arabic{figure}\alph{subfigure}}
\setcounter{subfigure}{1}
\begin{figure}
\epsscale{0.5}
\hspace{2cm}  \vspace{2cm}  \includegraphics[width=11cm,angle=0]{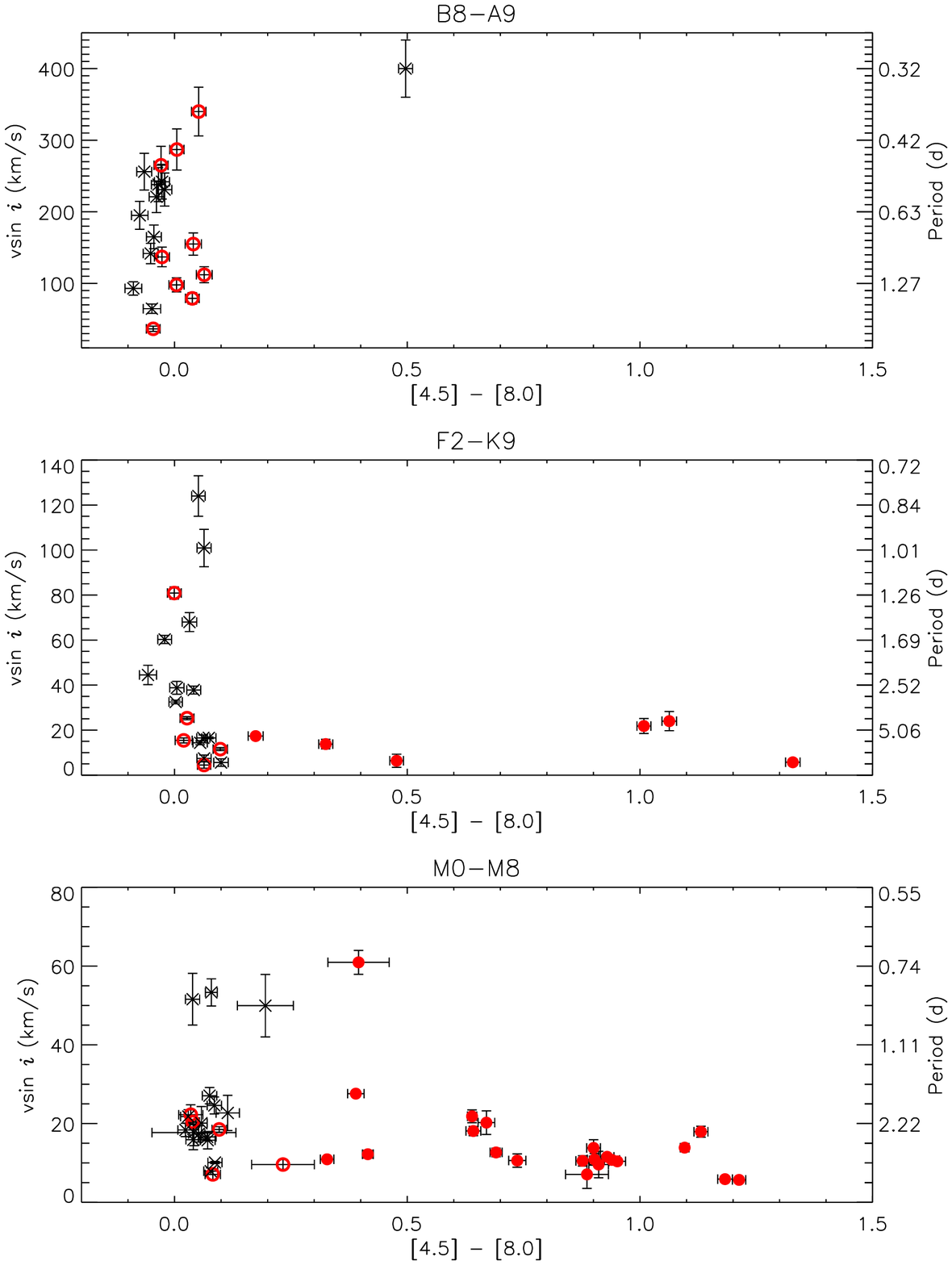} \vspace{-2cm}
\caption[f6a.ps]{a) Projected rotational velocity, $v$sin$i$, as a function of [4.5]$-$[8.0] color for
B8-A9 (top panel), F2--K9 (middle panel), and M0-M8 (bottom panel) sources in Upper Scorpius.
Symbols are as in Figure 4. To facilitate comparison with rotation period surveys, the mean velocities
for fiducial rotation periods are plotted on the right ordinate, assuming nominal stellar radii for 
each spectral type
or mass bin (2.5 R$_{\odot}$, 2.0 R$_{\odot}$, and 0.88 R$_{\odot}$, for B8--A9, F2--K9, and M0--M8
type sources, respectively). Among the early-type (B8-A9) stars, the rotational velocities of
the debris-disk and non-disk populations are indistinguishable. Considering the late-type (F2--M8)
stars and brown dwarfs, however, we find that the disk-bearing sources are in general displaced
toward lower $v$sin$i$ relative to their non-excess counterparts. There also appear to be few
slowly rotating, non-excess sources relative to younger star forming regions.
\label{f6a}}
\end{figure}
\clearpage

\addtocounter{figure}{-1}
\addtocounter{subfigure}{1}
\clearpage
\begin{figure}
\epsscale{0.5}
\hspace{2cm}  \vspace{2cm}  \includegraphics[width=11cm,angle=0]{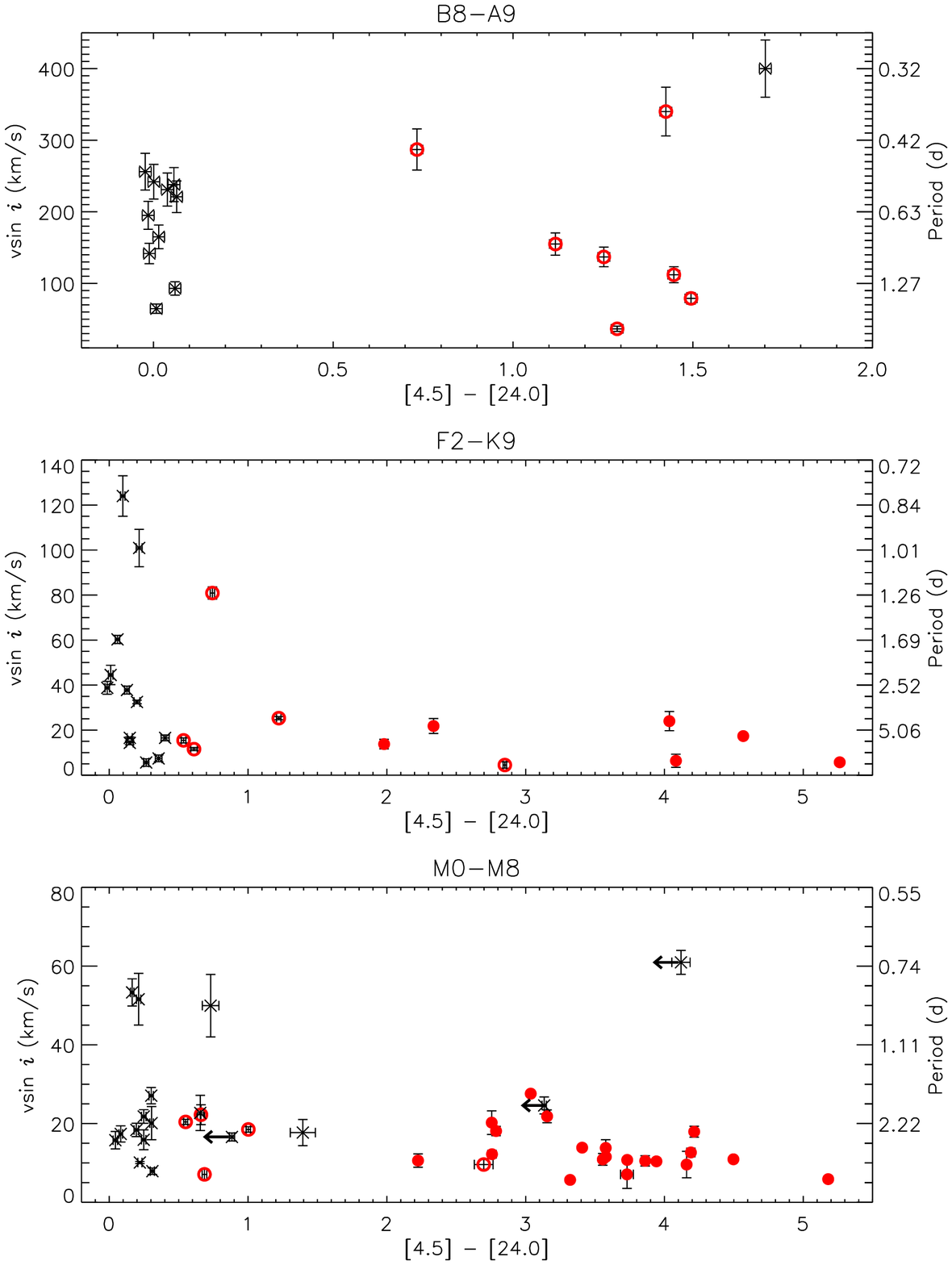}  \vspace{-2cm}    
\caption[f6b.ps]{b) Projected rotational velocity, $v$sin$i$, as a function of [4.5]$-$[24.0] color for
B8-A9 (top panel), F2--K9 (middle panel), and M0-M8 (bottom panel) sources in Upper Scorpius.
Symbols are as in Figure 4.
The MIPS 24 $\mu$m photometry for J161052.4$-$193734 is contaminated by bright nebulosity and
that for ScoPMS 17 by a nearby star. Their [4.5]$-$[24.0] colors are represented by upper limits.
SCH16235158$-$23172740 was not detected by MIPS and is also represented by a 3$\sigma$ upper limit.
\label{f6b}}
\end{figure}
\clearpage

\clearpage
\begin{figure}
\epsscale{0.65}
\hspace{2cm}  \vspace{2cm}  \includegraphics[width=11cm,angle=0]{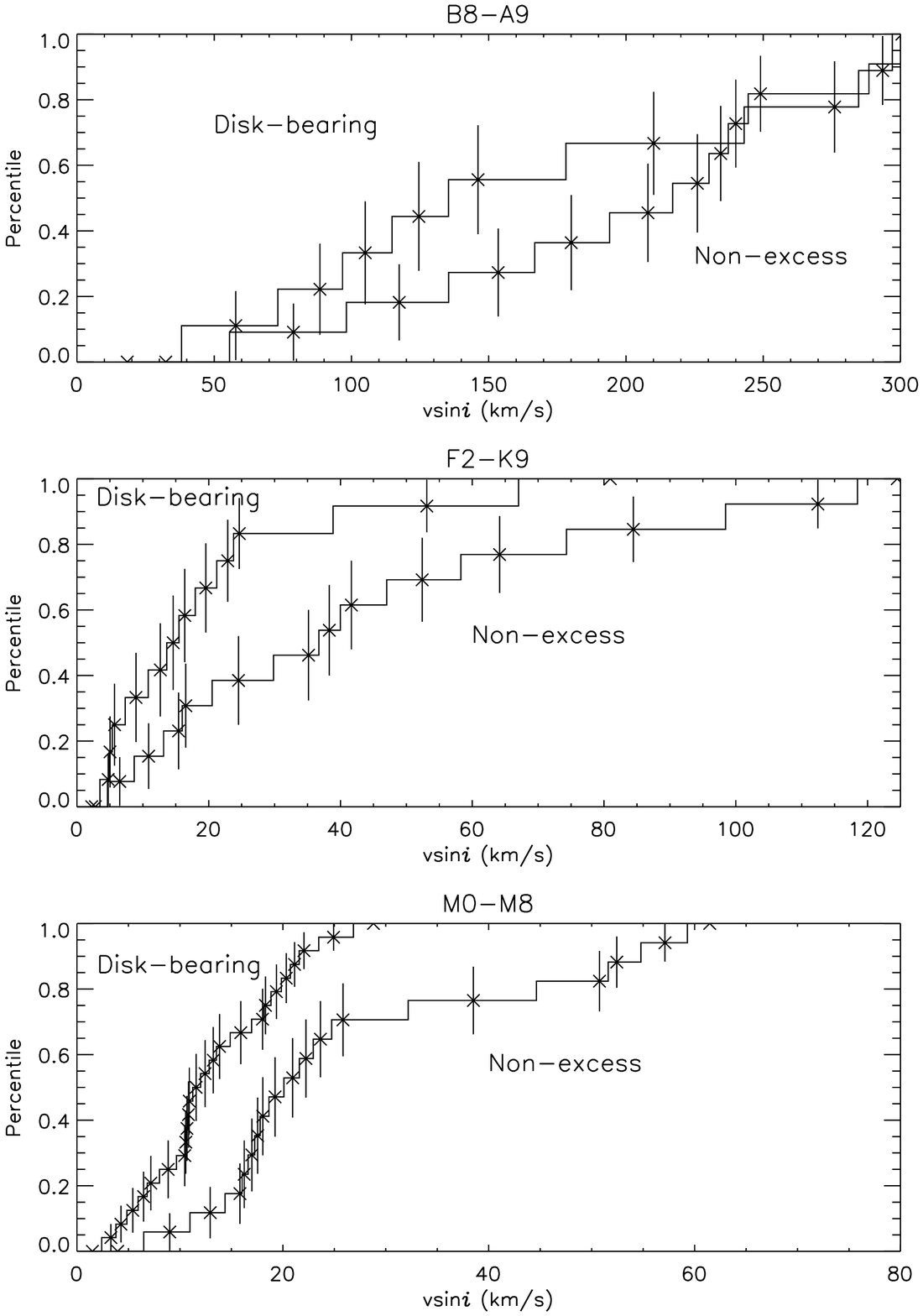}  \vspace{-2cm}    
\caption[f7.ps]{The cumulative distribution functions (CDF) for the disk-bearing and non-excess
stars and brown dwarfs in Upper Scorpius, by spectral type or mass bin: B8-A9 (top panel), F2--K9
(middle panel), and M0-M8 (lower panel). The CDFs for the late-type, disk-bearing stars and brown
dwarfs are substantially different from their non-disk counterparts, such that disk-bearing sources
are rotating more slowly on average.
\label{f7}}
\end{figure}
\clearpage

\clearpage
\begin{figure}
\epsscale{0.65}
\hspace{2cm}  \vspace{2cm}  \includegraphics[width=11cm,angle=0]{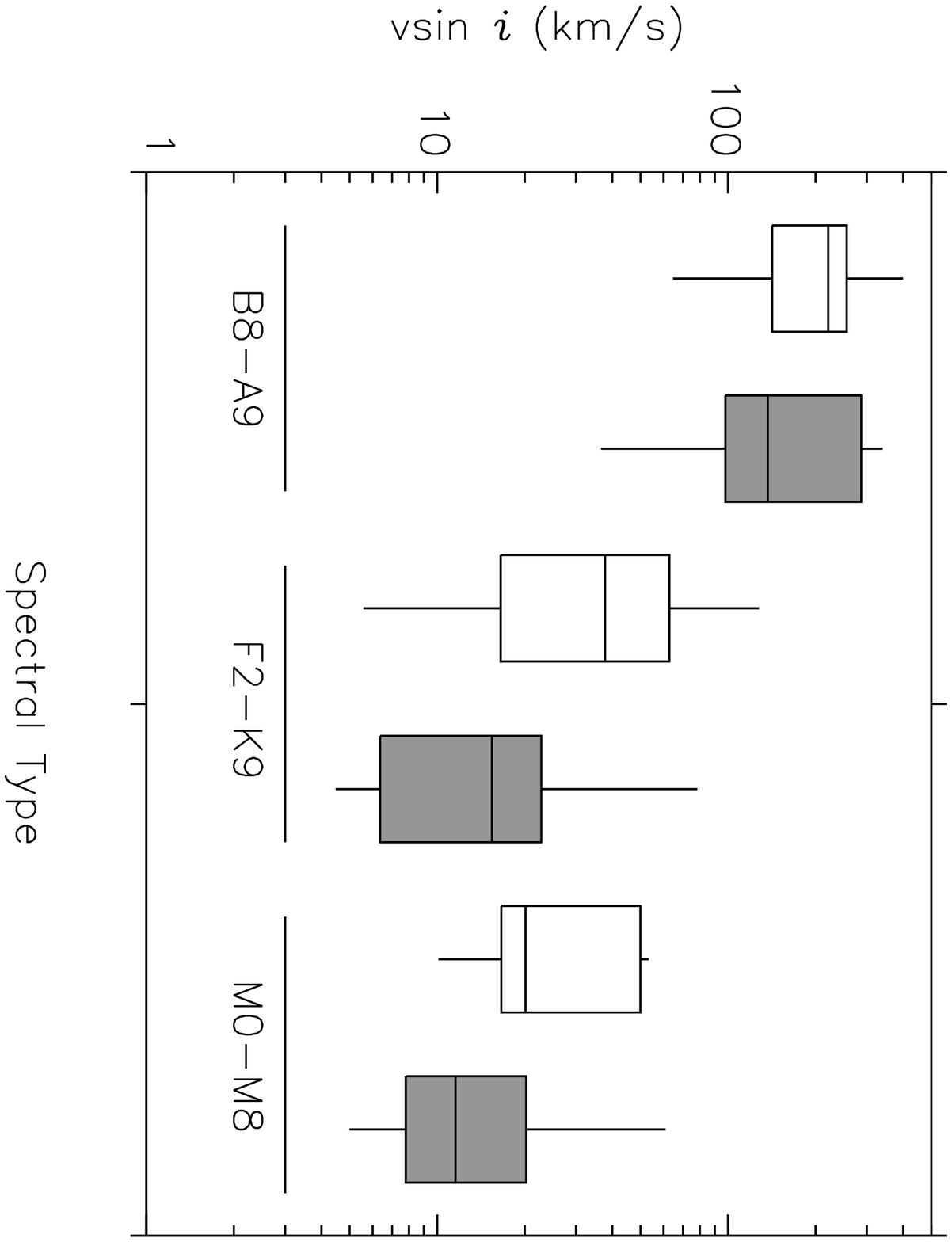}  \vspace{-2cm}    
\caption[f8.ps]{Boxplots of projected rotational velocities for disk-bearing and non-excess sources in
Upper Scorpius, grouped by spectral type. Shaded boxes represent stars with primordial or debris disks,
while clear boxes denote non-excess sources. The central rectangles represent the first and third quartiles
of each bin and the ``whiskers'' span the range of $v$sin$i$ values. The median of the distributions are
marked by the solid horizontal line within each box. We find statistically significant differences between
the rotational velocities for the disk-bearing and non-excess M-dwarfs in the association.
\label{f8}}
\end{figure}
\clearpage

\clearpage
\begin{figure}
\epsscale{0.65}
\hspace{2cm}  \vspace{2cm}  \includegraphics[width=11cm,angle=0]{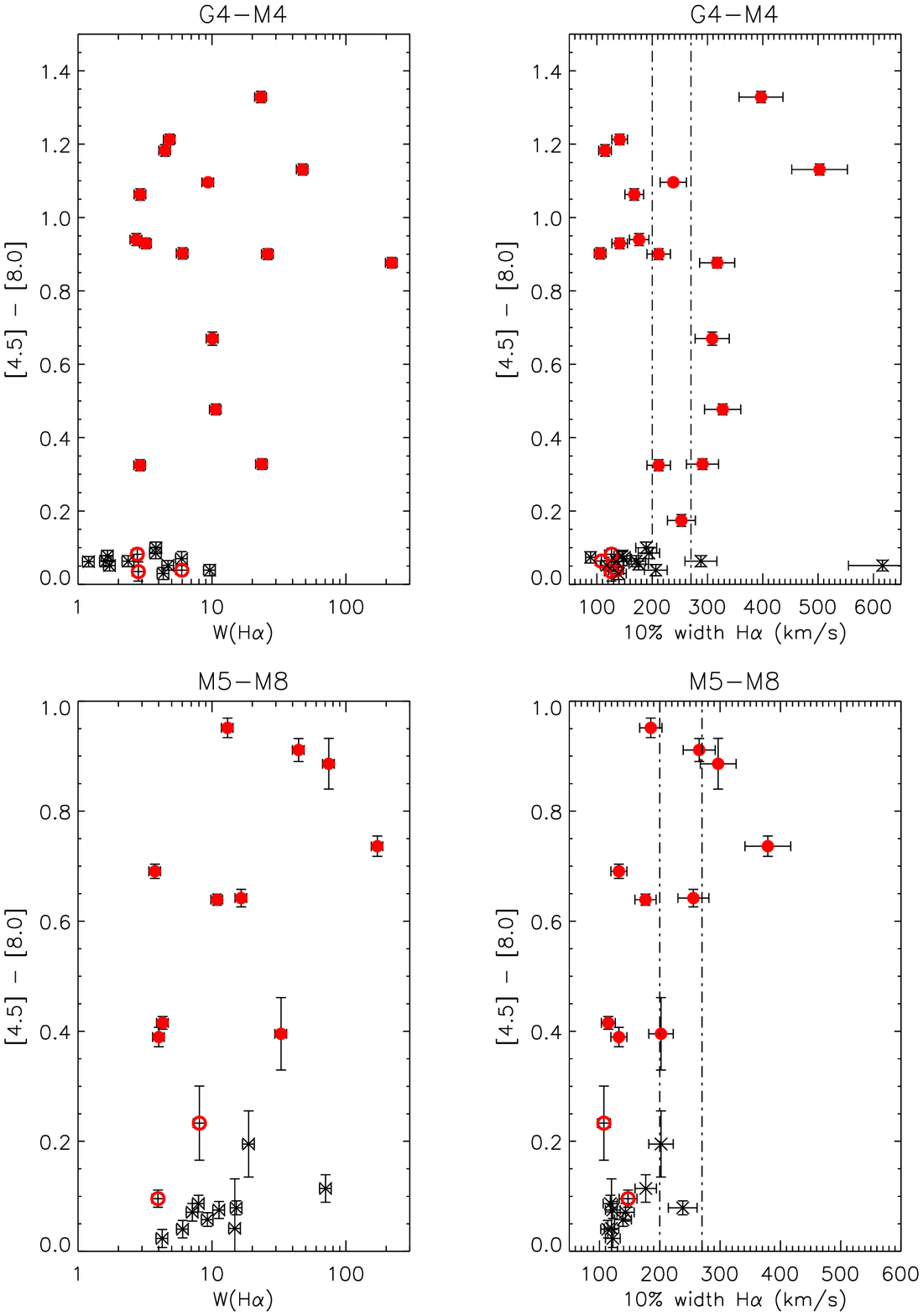}  \vspace{-2cm}    
\caption[f9.ps]{{\it Spitzer} IRAC [4.5]$-$[8.0] color plotted as a function of W(H$\alpha$) (left panels)
and H$\alpha$ 10\% velocity width (right panels) for the G4--M4 type stars in the Upper Scorpius spectroscopic 
sample (upper panels) and M5--M8 type sources (lower panels). Symbols are as in Figure 4. The vertical
dashed lines in the H$\alpha$ velocity width panels represent the 200 and 270 km s$^{-1}$ velocity widths 
associated with accretion from Jayawardhana et al. (2003) for sources M5 and later and White and Basri (2003)
for earlier spectral types. The errors shown represent the 1$\sigma$ photometric uncertainties and an assumed
uncertainty of 10\% for the measured W(H$\alpha$) values and for the H$\alpha$ velocity widths. The G4-type,
non-excess source HD 142987 (H$\alpha$ 10\% velocity width $\sim$600 km s$^{-1}$ is a rapid rotator and possible 
spectroscopic binary.
\label{f9}}
\end{figure}
\clearpage

\clearpage
\begin{figure}
\epsscale{0.65}
\hspace{2cm}  \vspace{2cm}  \includegraphics[width=11cm,angle=0]{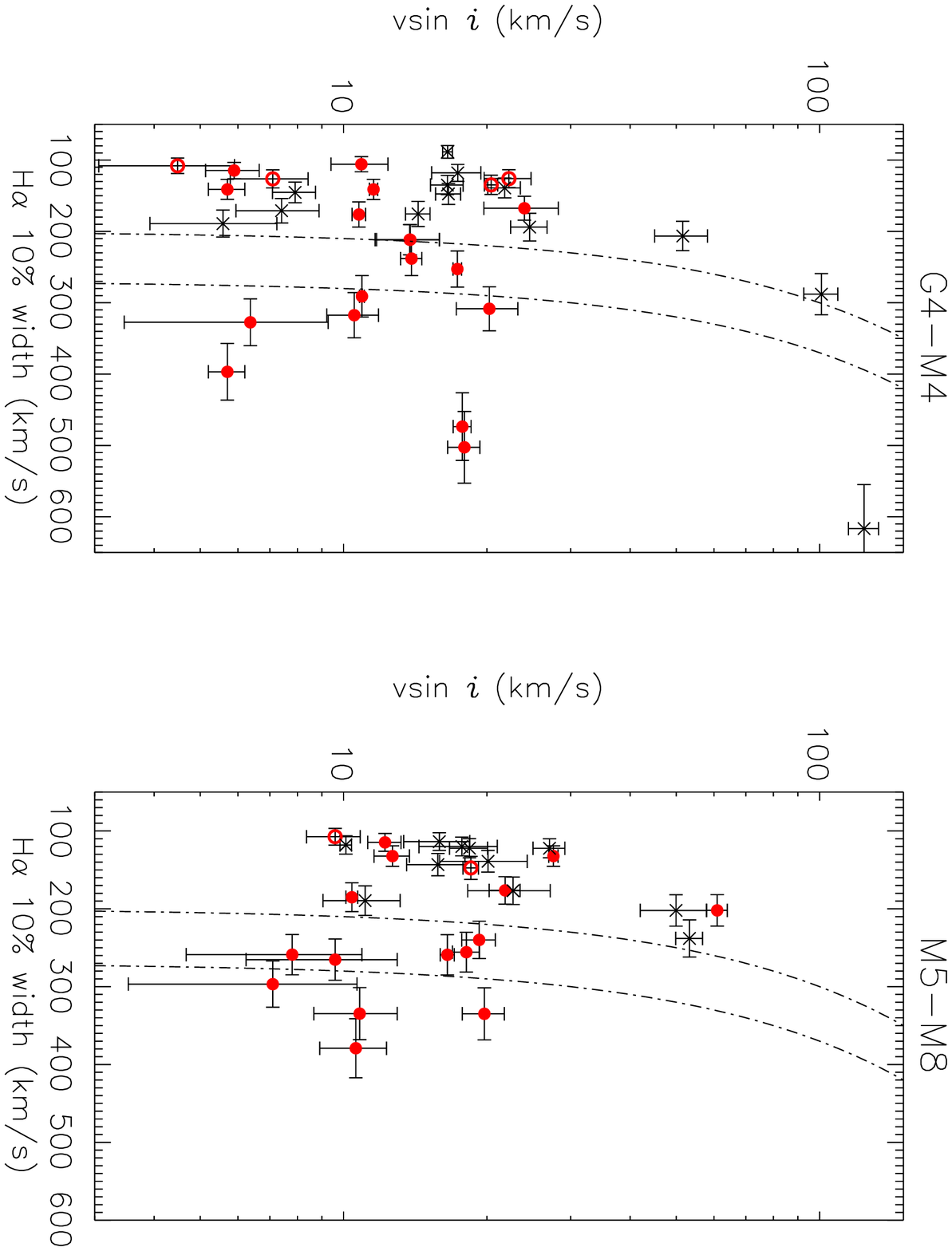}  \vspace{-2cm}    
\caption[f10.ps]{Projected rotational velocity plotted as a function of H$\alpha$ 10\% velocity width
for the G4--M4 type stars in the Upper Scorpius sample (left panel) and for M5--M8 type sources 
(right panel). The curved dashed lines represent the intrinsic contributions of rotational line
broadening to the 200 km s$^{-1}$ and the 270 km s$^{-1}$ velocity width criteria for accretion
from Jayawardhana et al. (2003) and White and Basri (2003), respectively.
\label{f10}}
\end{figure}
\clearpage

\clearpage
\begin{figure}
\epsscale{0.65}
\hspace{2cm}  \vspace{2cm}  \includegraphics[width=11cm,angle=0]{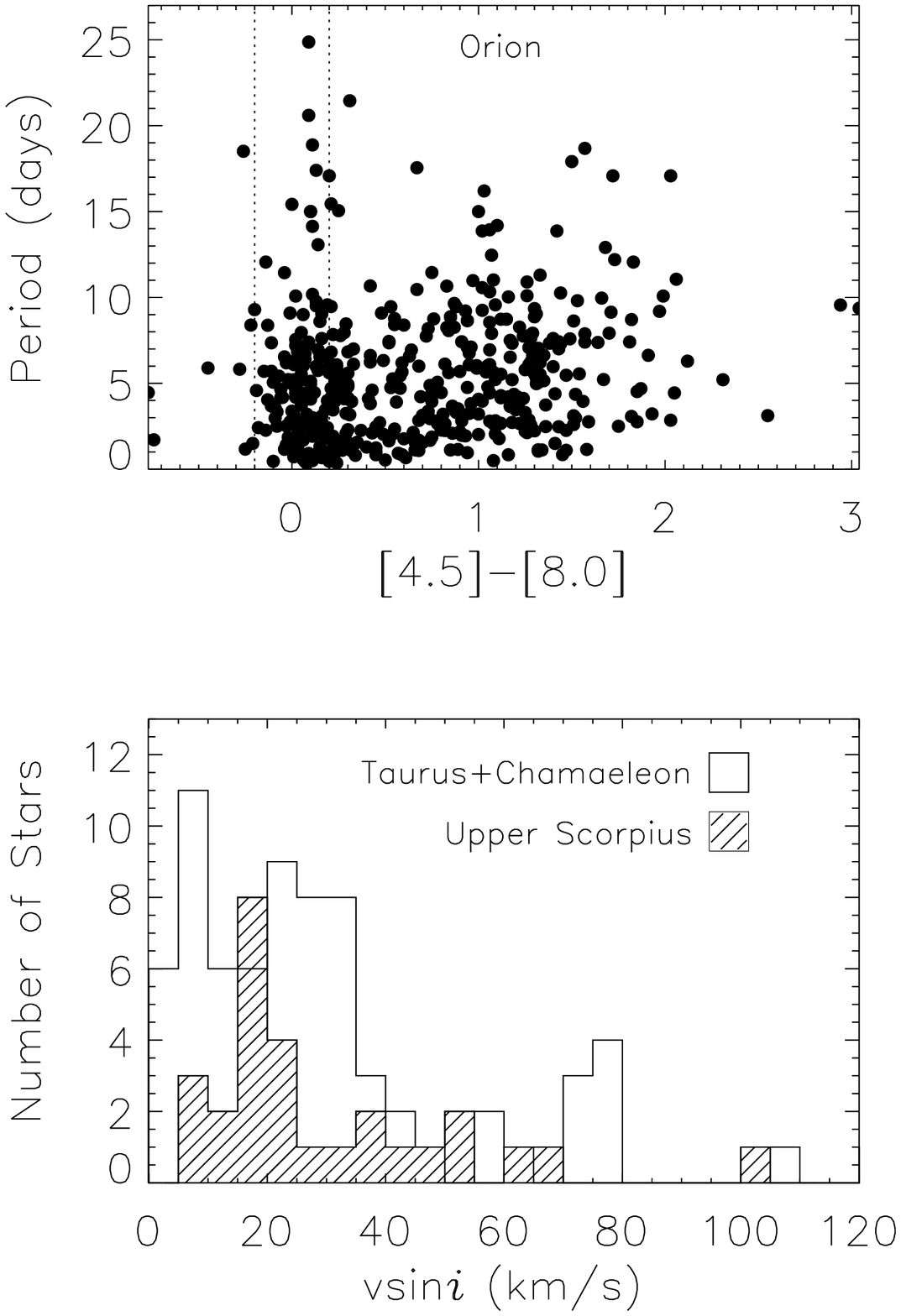}  \vspace{-2cm}    
\caption[f11.ps]{(upper panel) Rotation periods for 464 sources in Orion from Rebull et al. (2006) 
plotted as a function of [4.5]$-$[8.0] color. The vertical dot-dashed lines represent the adopted
color boundaries for non-excess (i.e. non-disk bearing) stars. The fraction of slowly-rotating,
non-excess sources in Orion having rotation periods of $>$5 days (corresponding to equatorial 
rotation velocities of $\le$15 km s$^{-1}$ for an assumed 1.5 $R_{\odot}$ star) is $\sim$45\% (63/140). 
(lower panel) Histograms
of measured $v$sin$i$ for late-type, non-excess sources in Upper Scorpius (cross-hatched region)
and Taurus-Auriga + Chamaeleon I (from Nguyen et al. 2009). The fraction of slowly rotating
($v$sin$i\lesssim15$ km s$^{-1}$), non-disk bearing sources in Upper Scorpius is $\sim$17\%
and that of Taurus and Chamaeleon I is $\sim$32\%.
\label{f11}}
\end{figure}
\clearpage

\begin{deluxetable}{lccccccccccc}
\tabletypesize{\tiny}
\tablenum{1}
\tablewidth{0pt}
\rotate
\tablecaption{Properties of the Upper Scorpius Spectroscopic Sample}
\tablehead{
\colhead{Source}  &  \colhead{Type} & \colhead{$V_{R}$\tablenotemark{a}} & \colhead{[4.5]\tablenotemark{b}} & \colhead{[8.0]\tablenotemark{b}} & \colhead{[24]\tablenotemark{b}} & \colhead{$v$sin$i$\tablenotemark{c}} & \colhead{W(H$\alpha$)\tablenotemark{d}} & \colhead{H$\alpha$ 10\% \tablenotemark{e}} & \colhead{Disk Status}  & \colhead{Epoch\tablenotemark{f}}  & \colhead{Comments}\\
                  &                 &    km s$^{-1}$    &     mJy                      &       mJy                       &        mJy                  &          km s$^{-1}$           &     \AA\                            &             km s$^{-1}$                &                         &            &     \\
}
\startdata
HIP 78207              & B8 & $+$1.54$\pm$2.23  & 4019.5  & 2268.5   & 766.4   & 400:, 38.2$\pm$2.0\tablenotemark{i}  &  $-$24.80 &  ... & Non-excess & H20070525 & Be, SB2 \\
HIP 79739              & B8 & $+$14.58$\pm$1.58  &  291.40 &   99.30  &  11.47  & 141.8$\pm$4.8 &  abs  &  ... & Non-excess  & M20080514 & Binary\tablenotemark{g}\\
HIP 76071              & B9 & $-$6.14$\pm$2.28  &  277.00 &   93.20  &  10.79  & 256$\pm$10    &  abs  &  ... & Non-excess  & M20080514  & Binary\tablenotemark{g}\\
HIP 77911              & B9 & $-$2.90$\pm$2.56  &  379.80 &  132.10  &  142.60 & 265$\pm$24    &  abs  &  ... & Debris      & H20070525  & Binary\tablenotemark{g}\\
HIP 78809              & B9 & $-$5.54$\pm$3.16 &  213.10 &   72.80  &   8.54  & 64.7$\pm$1.7  &  abs  &  ... & Non-excess  & M20080514  & \\ 
HIP 79410              & B9 & $-$6.15$\pm$2.57  &  279.10 &  104.60  &  41.25  & 340$\pm$19    &  abs  &  ... & Debris      & H20070524  & \\
HIP 79439              & B9 & $-$5.56$\pm$0.21  &  305.40 &  109.60  &  23.86  & 287$\pm$26    &  abs  &  ... & Debris      & H20070524  & \\
HIP 79785              & B9 & $-$6.37$\pm$0.20  &  443.90 &  153.60  &  18.61  & 238$\pm$13    &  abs  &  ... & Non-excess  & M20080514  & \\
HIP 80024              & B9 & $-$3.96$\pm$1.41  &  368.20 &  126.10  &  48.03  & 36.6$\pm$1.5  &  abs  &  ... & Debris      & H20070524  & \\
HIP 80493              & B9 & $-$7.43$\pm$1.92  &  285.20 &   97.80  &  11.50  & 165.1$\pm$6.2 &  abs  &  ... & Non-excess  & M20080514  & \\
HIP 76310              & A0 & $-$2.54$\pm$0.60  &  247.60 &   88.80  & 165.40  & 97.9$\pm$2.2  &  abs  &  ... & Debris      & H20070524  & \\
HIP 78847              & A0 & $-$23.00$\pm$3.0  &  261.70 &   91.20  &  10.42  & 242$\pm$14    &  abs  &  ... & Non-excess  & M20080514  & \\
HIP 79124              & A0 & $-$18.14$\pm$1.89 &  299.40 &  103.20  &  12.64  & 221$\pm$11    &  abs  &  ... & Non-excess  & M20080514  & Binary\tablenotemark{g}\\
HIP 79156              & A0 & $-$3.73$\pm$1.91  &  203.10 &   75.30  &  22.62  & 155.2$\pm$5.0 &  abs  &  ... & Debris      & H20070524  & Binary\tablenotemark{g}\\
HIP 79860              & A0 & $-$16.85$\pm$1.83 &  127.60 &   44.70  &   5.26  & 231$\pm$16    &  abs  &  ... & Non-excess  & M20080514  & \\
HIP 79878              & A0 & $-$3.43$\pm$0.57  &  268.90 &   93.70  &  33.90  & 137$\pm$5     &  abs  &  ... & Debris      & H20070524  & \\
HIP 78963              & A9 & $-$50.69$\pm$1.66 &  225.50 &   75.20  &   8.85  & 195$\pm$12    &  abs  &  ... & Non-excess  & M20080514  & \\
HIP 78996              & A9 & $-$7.88$\pm$2.00  &  237.80 &   90.10  &  35.87  & 112.1$\pm$10.1&  abs  &  ... & Debris      & H20070524  & \\
HIP 80088              & A9 & $-$7.52$\pm$2.22  &  143.50 &   53.10  &  22.62  & 79.1$\pm$4.7  &  abs  &  ... & Debris      & H20070524  & \\
HIP 80130              & A9 & $-$50.92$\pm$1.29 &  215.00 &   70.80  &   9.04  & 93.0$\pm$2.8  &  abs  &  ... & Non-excess  & M20080514  & \\
HIP 79643              & F2 & $-$1.97$\pm$3.08  &  117.90 &   42.10  &   9.30  & 80.9$\pm$2.7  &  abs  &  ... & Debris      & H20110424 &\\
HIP 82319              & F3 & $-$18.27$\pm$1.28 &  133.30 &   46.70  &    5.60 & 60.3$\pm$1.8  &  abs  &  ... & Non-excess  & M20080514  & \\
HIP 80896              & F3 & +3.33$\pm$2.67    &  201.50 &   68.30  &    8.09 & 44.5$\pm$4.3  &  abs  &  ... & Non-excess  & M20080514  & Binary\tablenotemark{g}\\
RXJ1550.9$-$2534       & F9 & $-$6.02$\pm$1.56  &  127.90 &   45.90  &    5.02 & 38.8$\pm$2.8  &  abs  & ...  & Non-excess  & M20080514  & \\
$[$PZ99$]$J160000.7$-$250941 & G0 & $-$1.33$\pm$0.30 & 56.70 & 20.30 &    2.71 & 32.5$\pm$0.8  &  abs  &  ... & Non-excess  & H20110424 &\\
$[$PZ99$]$J155812.7$-$232835 & G2 & $-$4.56$\pm$0.21 & 117.70& 43.10 &   14.42 & 25.3$\pm$0.7  &  abs  &  ... & Debris      & H20110319 &\\ 
HIP 79462              & G2 & $-$4.99$\pm$1.16\tablenotemark{h} &  232.00 &   84.40  &   15.11 & 15.43$\pm$1.11 & abs & ... & Debris & H20110424 & SB2 \\
HD 142361              & G3 V &$-$5.84$\pm$1.62 & 292.70  &  107.70  &    3.52 & 68.0$\pm$4.2  &  abs  &  ... & Non-excess  & H20110424 &  \\
HD 142987              & G4 & $-$6.43$\pm$2.09  & 181.30  &   67.90  &    7.89 & 124.0$\pm$9.0 & $-$1.72 & 616.5 & Non-excess & H20110424 & Rapid rotator\\ 
RXJ16036-2245          & G9 & $-$10.41$\pm$0.18 &   89.20 &   33.10  &    3.99 & 37.8$\pm$1.8  &  abs  & ...    & Non-excess & M20080514 &  \\
$[$PZ99$]$J161411.0$-$230536 & K0 & $-$6.31$\pm$1.41  &  401.90 &  363.50  &  137.49 & 21.8$\pm$3.3 & abs       & ...  & Primordial & H20060616 & sub-arcsec binary\\
$[$PZ99$]$J160421.7$-$213028 & K2 & $-$6.90$\pm$0.24  &   62.70 &   26.30  &  167.50 & 17.3$\pm$0.4 & $-$0.27   & 252.7 & Primordial & H20060616 & Possible accretor \\
$[$PZ99$]$J155847.8$-$175800 & K3 & $-$7.17$\pm$0.10  &   93.20 &   35.30  &  2.18   & 4.48$\pm$1.42 & $-$0.76  & 108.0 & Debris & H20110319 &  \\
$[$PZ99$]$J160251.2$-$240156 & K4 & $-$6.53$\pm$0.20  &   50.50 &   19.10  &  2.91   & 16.5$\pm$1.3 & $-$1.20   & 135.0 & Non-excess & H20110424 & \\
ScoPMS 45              & K5 IV    & $-$7.86$\pm$0.10  &   78.50 &   30.70  &  5.48   & 11.54$\pm$0.70 & abs & ... & Debris & H20110424 & \\
$[$PZ99$]$J160357.6$-$203105 & K5 & $-$6.66$\pm$0.88  &  191.90 &  106.40  &  328.00 &  6.37$\pm$2.91 & $-$10.65  & 327.38 & Primordial & H20060616 & Accretor \\
$[$PZ99$]$J160856.7$-$203346 & K5 & $-$9.18$\pm$0.20  &   73.50 &   28.10  &    3.35 & 16.54$\pm$0.39 & $-$0.10   &  88.46 & Non-excess & M20080514 &  \\
J160643.8-190805       & K6 & $-$5.43$\pm$0.56        &   52.30 &   25.20  &   12.89 & 13.77$\pm$2.12 & $-$2.90   & 211.61 & Primordial & H20060616 & sub-arcsec binary \\
$[$PZ99$]$J160042.8$-$212737 & K7 & +1.99$\pm$7.08    &   53.60 &   20.30  &    2.60 & 100.9$\pm$8.3  & $-$2.37   & 287.98 & Non-excess & H20110424 & Rapid rotator\\ 
$[$PZ99$]$J160239.1$-$254208 & K7 & $-$3.60$\pm$0.15  &   46.30 &   17.40  &    2.11 & 14.34$\pm$0.85 & $-$0.74   & 175.50 & Non-excess & H20110424 &  \\
J161031.9$-$191305           & K7 & $-$6.91$\pm$0.27  &   60.00 &   23.50  &    3.05 &  5.58$\pm$1.66 & $-$3.82   & 189.02 & Non-excess & H20110424 &  \\
J160801.4$-$202741           & K8 & $-$7.08$\pm$0.36  &   40.40 &   15.30  &    2.23 &  7.41$\pm$1.47 & $-$1.61   & 170.99 & Non-excess & H20110424 &  \\
J160823.2$-$193001       & K9 & $+$6.18$\pm$2.68  &   47.00 &   44.70  &   76.90 &  23.98$\pm$4.27 & $-$2.91  & 167.51 & Primordial & H20060616 & SB2(?)  \\ 
J160900.7$-$190852       & K9 & $-$7.30$\pm$0.23  &   56.40 &   68.50  &  285.80 &  $\le$5.7        & $-$23.17 & 396.75 & Primordial & H20060616 & Accretor \\
$[$PZ99$]$J160831.4$-$180241 & M0 & $-$9.73$\pm$0.12  &   54.20 &   20.80  &    2.87 &  7.91$\pm$0.82 & $-$1.66   & 145.34 & Non-excess & M20080514 & \\
J161420.2$-$190648       & M0 & $-$6.77$\pm$1.77  &  606.50 &  614.10  & 1170.00 &  17.94$\pm$1.39 & $-$47.42 & 502.56 & Primordial & H20060616 & Accretor \\
ScoPMS 31              & M0.5 & $-$4.53$\pm$0.27 & 133.70 &   64.60  &  334.60 &  10.93$\pm$0.13 & $-$23.53 & 290.93 & Primordial & H20060616 & Accretor, sub-arcsec binary \\
ScoPMS 17              & M1 & $-$3.70$\pm$0.75  &   60.80 &   23.50  &   43.36 &  24.60$\pm$2.16 & $-$3.80  & 193.97 & Non-excess & H20060616 & \\
J160954.4$-$190654       & M1 & $-$6.87$\pm$0.23  &   31.40 &   12.10  &    2.35 &  $\le$7.1         & $-$2.77  & 126.38 & Debris     & M20080513 & \\
J161115.3$-$175721       & M1 & $-$7.23$\pm$0.17  &   86.00 &   93.90  &   72.82 &  $\le$5.7         & $-$4.82  & 141.08 & Primordial & H20060616 & \\
J160341.8$-$200557       & M2 & $-$3.71$\pm$1.00\tablenotemark{h}   &   35.00 &   13.00  &    1.76 &  ...            & $-$1.70  & 198.00 & Non-excess & M20080513 & SB2 \\
J160357.9$-$194210       & M2 & $-$3.35$\pm$0.16  &   23.90 &   20.10  &   25.65 &  11.56$\pm$0.23 & $-$3.20  & 141.10 & Primordial & H20060616 & \\
J160545.4$-$202308       & M2 &    ...            &   21.20 &   18.00  &   26.22 &  10.77$\pm$0.35 & $-$2.72  & 176.30 & Primordial & H20060616 & SB2, Visual binary  \\
J155829.8$-$231007       & M3 & $-$4.37$\pm$3.03  &   17.60 &   14.10  &   24.53 &  10.53$\pm$1.31 & $-$219.51& 317.41 & Primordial & H20070524 & Accretor \\
J160953.6$-$175446       & M3 & $-$5.34$\pm$1.28  &    8.55 &    7.00  &    9.17 &  13.81$\pm$2.08 & $-$26.03 & 211.61 & Primordial & H20070524 & \\
J161052.4$-$193734       & M3 & $-$7.94$\pm$1.36  &   12.70 &    4.84  &    1.14 &  16.61$\pm$1.00 & $-$5.94  & 147.44 & Non-excess & M20080513 & \\
J155624.8$-$222555       & M4 & $-$6.32$\pm$0.86  &   22.80 &   18.70  &   23.98 &  10.90$\pm$1.49 & $-$6.02  & 105.80 & Primordial & H20070524 & \\
J155706.4$-$220606       & M4 & $-$6.02$\pm$1.80  &   15.10 &   10.00  &    7.60 &  20.24$\pm$2.99 & $-$10.08 & 308.38 & Primordial & H20070525 & Accretor \\
J155729.9$-$225843       & M4 & $-$1.06$\pm$1.03  &    6.65 &    6.52  &    6.10 &  13.89$\pm$0.71 & $-$9.35  & 238.07 & Primordial & H20070525 &  \\
J155918.4$-$221042       & M4 & $-$7.93$\pm$0.77  &   24.20 &    8.96  &    1.17 &  51.58$\pm$6.55 & $-$9.62  & 206.41 & Non-excess & M20080513 & SB2(?)\\
J160439.1$-$194245       & M4 & $-$3.67$\pm$2.26  &   11.90 &    4.39  &    0.87 &  22.25$\pm$2.51 & $-$2.82  & 126.0 & Debris     & H20110319 &  \\
J160708.7$-$192733       & M4 & $-$1.97$\pm$0.91  &    8.48 &    3.14  &    0.56 &  20.42$\pm$0.66 & $-$5.95  & 134.77 & Debris     & M20080516 &  \\
J160801.5$-$192733       & M4 & $-$4.44$\pm$0.50  &   31.60 &   11.60  &    1.58 &  21.80$\pm$1.72 & $-$4.33  & 139.00 & Non-excess & M20080515 & \\
J160959.4$-$180009       & M4 & $-$6.28$\pm$0.80  &   24.20 &   25.70  &  113.60 &   5.89$\pm$0.76 & $-$4.45  & 114.63 & Primordial & H20070525 & \\
J161026.4$-$193950       & M4 & $-$7.43$\pm$1.69  &   19.80 &    7.42  &    0.85 &  17.37$\pm$2.06 & $-$4.70  & 117.97 & Non-excess & M20080515 & \\
SCH16014156$-$21113855   & M4 & $-$12.81$\pm$1.00 &     ... &    ...   &     ... &  17.76$\pm$0.77 & $-$95.95 & 473.57 & ...        & M20080516 & Accretor \\
J160159.7$-$195219       & M5 & $-$5.32$\pm$1.57  &    6.77 &    2.62  &    0.33 &  10.11$\pm$0.28 & $-$7.93  & 117.93 & Non-excess & M20080513 &  \\
J160210.9$-$200749       & M5 & $-$4.28$\pm$0.76  &    7.00 &    2.67  &    0.29 &  15.78$\pm$2.21 & $-$7.13  & 143.22 & Non-excess & M20080515 &  \\
J160449.9$-$203835       & M5 & $-$3.06$\pm$2.08  &   11.20 &    4.22  &    0.59 &  20.11$\pm$4.20 & $-$9.25  & 138.99 & Non-excess & M20080515 &  \\
J160525.5$-$203539       & M5 & $-$3.26$\pm$1.75  &   11.80 &    6.18  &    5.95 &  12.22$\pm$0.98 & $-$4.28  & 114.63 & Primordial & H20070525 & \\
J160531.5$-$192623       & M5 & $-$5.22$\pm$1.17  &    6.66 &    2.55  &    0.35 &  27.08$\pm$2.08 & $-$11.28 & 122.19 & Non-excess & M20080513 &  \\
J160532.1$-$193315       & M5 & $-$3.36$\pm$1.58  &   12.70 &    8.94  &    3.92 &  10.61$\pm$1.70 & $-$171.67& 379.12 & Primordial & H20070525 & Accretor \\
J160600.6$-$195711       & M5 & $-$5.44$\pm$1.80  &   21.90 &   11.20  &   14.28 &  27.60$\pm$0.62 & $-$4.01  & 132.25 & Primordial & H20070524 & \\
J160611.9$-$193532       & M5 & $-$7.38$\pm$5.84  &   10.80 &    4.15  &    0.50 &  53.30$\pm$3.45 & $-$15.21 & 238.04 & Non-excess & H20070525 & Rapid rotator\\
J160622.8$-$201124       & M5 & $-$5.09$\pm$0.68  &   10.40 &    7.02  &   19.63 &  12.67$\pm$1.08 & $-$3.75  & 132.25 & Primordial & H20070524 & \\
J160702.1$-$201938       & M5 & $-$4.83$\pm$2.72  &   11.90 &    7.66  &    8.65 &  21.85$\pm$1.63 & $-$10.90 & 176.32 & Primordial & H20070524 & Visual binary\\
J160802.4$-$202233       & M5 & $-$5.84$\pm$1.30  &   15.40 &    5.71  &    0.77 &  15.89$\pm$2.52 & $-$6.04  & 113.71 & Non-excess & M20080515 & \\
J160827.5$-$194904       & M5 & $-$6.49$\pm$2.11  &   17.20 &   11.10  &    8.91 &  18.11$\pm$1.18 & $-$16.51 & 255.69 & Primordial & H20070524 & Accretor \\
J160900.0$-$190836       & M5 & $-$8.34$\pm$1.48  &   14.10 &   12.10  &   21.17 &  10.41$\pm$0.30 & $-$13.07 & 185.13 & Primordial & H20070524 & \\
J160915.8$-$193706       & M5 & $-$4.10$\pm$1.11  &    6.93 &    2.53  &    0.33 &  18.38$\pm$1.69 & $-$4.25  & 122.14 & Non-excess & M20080515 & \\
J161011.0$-$194603       & M5 & $-$8.52$\pm$0.87  &    7.79 &    3.04  &    0.78 &  18.51$\pm$0.69 & $-$3.95  & 147.44 & Debris     & M20080515 & \\
SCH16033470$-$18293060   & M5 & $-$7.34$\pm$2.37  &    ...  &     ...  &    ...  &  19.76$\pm$2.00 & $-$25.06 & 334.87 & ...        & M20080516 & Accretor \\
SCH16081081$-$22294303   & M5 & $-$10.93$\pm$2.05 &    ...  &     ...  &    ...  &  16.52$\pm$4.05 & $-$30.54 & 259.05 & ...        & M20080516 & Accretor \\
SCH16150524$-$24593500   & M5 & $-$16.22$\pm$2.12 &    ...  &    ...   &     ... &   7.80$\pm$3.13 & $-$21.83 & 258.88 & ...        & M20080516 & Accretor \\
SCH16092137$-$21393452   & M5.5 & $-$12.63$\pm$0.3&    ...  &     ...  &    ...  &  11.10$\pm$2.05 & $-$22.79 & 189.44 & ...        & M20080516 & \\
SCH16305349$-$24245439   & M5.5 & $-$8.87$\pm$0.46&    ...  &     ...  &     ... &  19.27$\pm$1.56 & $-$30.31 & 239.94 & ...        & M20080516 & Accretor \\
SCH16093018$-$20595409   & M6 & $-$5.08$\pm$0.60  &    1.94 &    0.859 &    0.93 &   $\le$9.6      & $-$8.05  & 107.36 & Debris (?) & M20080513 & \\
SCH16103876$-$18292353   & M6 & $-$6.83$\pm$0.82  &    3.70 &    2.99  &    4.57 &   $\le$7.1      & $-$74.44 & 296.75 & Primordial & M20080514 & Accretor \\
SCH16200756$-$23591522   & M6 & $-$10.71$\pm$5.06 &    4.49 &    1.92  &    0.35 &  49.94$\pm$7.93 & $-$18.78 & 202.06 & Non-excess & M20080513 & SB2(?)\\
SCH16202127$-$21202923   & M6 & $-$5.92$\pm$0.90  &    3.34 &    1.24  &    0.48 &  17.72$\pm$3.32 & $-$14.81 & 119.96 & Non-excess & M20080514 & \\
SCH16263026$-$23365552   & M6 & $-$12.15$\pm$1.20 &    8.09 &    6.69  &   14.84 &   $\le$9.6       & $-$44.31 & 265.28 & Primordial & M20080515 & Accretor \\
SCH16284706$-$24281413   & M6 & $-$10.56$\pm$0.50 &    ...  &     ...  &     ... &  10.81$\pm$2.15 & $-$189.17& 334.67 & ...        & M20080516 & Accretor \\
SCH16224384$-$19510575   & M8 & $-$7.87$\pm$1.28  &   11.64 &    4.62  &    0.85 &  22.70$\pm$4.47 & $-$70.52 & 176.77 & Non-excess & M20080515 & SB2(?)\\
SCH16235158$-$23172740   & M8 & $-$7.47$\pm$2.45  &    3.52 &    1.81  & $<$6.22 &  60.94$\pm$3.03 & $-$32.75 & 202.02 & Primordial & M20080514 & Rapid rotator\\
\enddata
\tablenotetext{a}{Heliocentric radial velocity.}
\tablenotetext{b}{IRAC and MIPS fuxes from Carpenter et al. (2006) and Carpenter et al. (2009), respectively.}
\tablenotetext{c}{Projected rotational velocities.}
\tablenotetext{d}{Equivalent width of H$\alpha$. Negative values imply emission.}
\tablenotetext{e}{Velocity width of H$\alpha$ emission profiles at 10\% of peak flux.}
\tablenotetext{f}{UT Date of observation (yyyymmdd) with prefix: H=HIRES, M=MIKE}
\tablenotetext{g}{Binary identified by Kouwenhoven et al. (2007).}
\tablenotetext{h}{Systemmic velocity reported for double-line spectroscopic binaries.}
\tablenotetext{i}{$v$sin$i$ provided for both primary and secondary components.}
\end{deluxetable}

\begin{deluxetable}{lcc}
\tabletypesize{\tiny}
\tablenum{2}
\tablewidth{0pt}
\rotate
\tablecaption{Radial and Rotational Velocity Measurements from the Literature}
\tablehead{
\colhead{Source} & \colhead{$V_{R}$\tablenotemark{a,b}}   &   \colhead{$v$sin$i$}\tablenotemark{b,c}\\
                 & (km s$^{-1}$)             &  (km s$^{-1}$)                 \\
}
\startdata
HIP 78207                       & $-$5.6 (1)                           &   400  (2)    \\
HIP 79739                       & $-$14.7 (3)                          &   160  (2)    \\
HIP 77911                       & +3 (ref 1); $-$7.4 (3)               &   300: (2)    \\
HIP 79410                       & $-$9.4 (3)                           &   180  (2)    \\
HIP 79439                       & $-$9.5 (3)                           &   160: (2)    \\
HIP 79785                       & $-$8 (1); $-$10.7 (3)                &   300: (2)    \\
HIP 80024                       & $-$9 (4); $-$7.9 (3)                 &  $\le$50 (2)  \\
HIP 76310                       & +4.4 (3)                             &   ...         \\
HD 142361                       & $-$3.39 (5); $-$3.9 (6)              &    56 (6)     \\
$[$PZ99$]$J155812.7$-$232835    & $-$7.0 (6)	                       &   ...         \\
$[$PZ99$]$J161411.0$-$230536    & $-$4.9 (6)	                       &   27.5 (6)    \\
RXJ16036-2245	                & $-$5.7 (6)	                       &   34.0 (6)    \\
ScoPMS 45        		& $-$7.7 (7)		               &   14 (7)      \\
ScoPMS 31        		& $-$3.7 (7) 		               &  $<$15 (7)   \\
ScoPMS 17        		& $-$3.5 (7)		               &   35 (7)      \\
SCH16224384$-$19510575  	& $-$10.2 (8) 		               &   25 (8) \\
SCH16235158$-$23172740  	& $-$6.0 (8)		               &   53 (8) \\
\enddata
\tablenotetext{a}{Heliocentric radial velocity.}
\tablenotetext{b}{References: 1) Evans (1967); 2) Slettebak (1968); 3) Sartori et al. (2003); 4) Wilson (1953); 5) White et al. (2007); 6) Torres et al. (2006); 7) Walter et al. (1994); 8) Rice et al. (2008).}
\tablenotetext{c}{Projected rotational velocity.}

\end{deluxetable}

\end{document}